\documentstyle[12pt,aasms4]{article}
\lefthead{et al.}
\righthead{Photometry of QSO 0957+561A, B}

\begin{document}

\title {
%BVRI photometry of QSO 0957+561A, B\\
$BVRI$ photometry of QSO 0957+561A, B: Observations, new reduction 
method and time delay\\
}

\author{Miquel Serra--Ricart$^1$, Alejandro Oscoz$^1$, 
Teresa Sanch\'\i s$^1$, Evencio Mediavilla$^1$, 
Luis Juli\'an Goicoechea$^2$, Javier Licandro 
$^{1, 3}$, David Alcalde$^1$, and Rodrigo Gil-Merino$^2$}
\affil{
(1)Instituto de Astrof\'\i sica de Canarias, E--38200, La Laguna, 
Tenerife, Spain
(2)Departamento de F\'\i sica Moderna, Universidad de 
Cantabria, E-39005, Santander, Cantabria, Spain \\
(3) Departamento de Astronom\'\i a, Facultad de Ciencias, Universidad 
de la Rep\'ublica, Montevideo, Uruguay
\\
Electronic mail:  mserra@ot.iac.es, aoscoz@ll.iac.es, 
tsanchis@ll.iac.es, emg@ll.iac.es, goicol@besaya.unican.es,
 jlicandr@ll.iac.es, dalcalde@ll.iac.es, rgil@ll.iac.es}

\begin{abstract}
CCD observations of the gravitational lens system Q0957+561A,B in 
the $BVRI$ bands are presented in this paper. The observations, 
taken with the 82 cm IAC-80 telescope, at Teide Observatory, Spain, 
were made from the beginning of 1996 February to 1998 July, as part 
of an on-going lens monitoring program. Accurate photometry was 
obtained by simultaneously fitting a stellar two-dimensional 
profile on each component by means of DAOPHOT software. This 
alternative method equals and even improves the results obtained 
with previous techniques. The final dataset is characterized by its 
high degree of homogeneity as it was obtained using the same 
telescope and instrumentation during a period of almost 3 years. 
The resulting delay, obtained with a new 
method, the $\delta^2$-test, is of 425 $\pm$ 4 days, 
slightly higher than the value previously accepted (417 days), but 
concordant with the results obtained by Pelt et al. (1996); Oscoz 
et al. (1997); Pijpers (1997); and Goicoechea et al. (1999).
\end{abstract}


\section{Introduction} \label{intro}

Since the discovery of the first gravitational lens, the Twin QSO
0957+561 (Walsh et al. 1979), the system has been subject to the
most rigorous attempt to measure the time delay between its 
components. The specially propitious configuration of QSO 0957+561, 
two images separated by $6\farcs1$ of a quasar ($z = 1.41$) lensed 
by a galaxy ($z = 0.36$) placed at the center of a cluster of 
galaxies, make it suitable for photometric monitorings. Although 
the time delay controversy has recently been solved, establishing 
a value for the time delay of $\sim 420$ days (Oscoz et al. 1997; 
Kundi\'c et al. 1997;  Pelt et al. 1998a), an on-going monitoring 
of the QSO components and comparison between the light-curves may 
yield important results, both for the study of the physical 
properties of quasars (Peterson 1993, Gould \& Miralda-Escud\'e 
1997) and for the detection of possible microlensing events
(Gott 1981, Pelt et al. 1998b). Moreover, and although it will not 
lead to substantial changes in the value of the Hubble constant, a 
secure statement of the time delay is crucial for microlensing 
studies.

The main requirement to obtain useful information from the 
light-curves of the two components is a high level of photometric 
accuracy. However,the QSO 0957+561 is a very complicated system for 
two main reasons: i) the proximity of the point-like QSO 
components; and ii) the extended light distribution of the 
underlying lensing galaxy. Moreover, the whole scenario presents 
an additional complication due to the large amount of available 
data to reduce and analyze (Vanderriest et al. 1989; 
Press et al. 1992; Schild \& Thomson 1995), hence automatic 
photometry codes become mandatory. Up to now, the 
only automated solution presented was developed by Colley \& 
Schild (1999), who used HST data to subtract the lens galaxy and 
estimate the level of cross talk between the QSO components by 
selecting reference stars. 

In this paper an alternative solution to this problem is presented:
PSF fitting by means of DAOPHOT software. To check the feasibility 
of this new technique, it was applied to a sample of simulated 
data. 

The dataset presented here is the result of 
three years monitoring, from 1996 February to 1998 July, 
a program which included 220 sessions of observation 
in the $R$ band, 62 in the $B$
band, 72 in the $V$ band and 68 in the $I$ band. The data acquisition 
and reduction processes are explained in detail in \S 2 and \S 3,
respectively. The software environment used for the different 
reduction and analysis processes was IRAF\footnote{IRAF is 
distributed by the National Optical Astronomy Observatories, which 
are operated by the Association of Universities for Research in 
Astronomy, Inc., under cooperative agreement with the National 
Science Foundation.} (Image Reduction and Analysis Facility, see 
http://www.noao.edu for more information), so any task or package 
referred to elsewhere is included in the IRAF environment. Section 4 
is devoted to presenting the observed light curves and in \S 5 we 
discuss on the time delay obtained from these data. Finally, a 
brief summary of the results is given in \S 6.

\section{Data Acquisition} \label{obser}

Lens monitoring was performed in three consecutive seasons, 
1996 February to June, 1996 October to 1997 July, and 1997 October 
to 1998 July (96, 97 and 98 season respectively hereby), using the 
CCD camera of the 82 cm IAC-80 telescope, sited at the Instituto 
de Astrof\'\i sica de Canarias' Teide Observatory (Tenerife, 
Canary Islands, Spain). A Thomson 1024$\times$1024 chip was used, 
offering a field of nearly 7$\farcm$5. Standard $BVRI$ broadband 
filters were used for the observations, corresponding fairly closely 
to the Landolt system (Landolt 1992). The IAC Time Allocation 
Committee awarded time for two kinds of observing runs: routine 
observation nights (RON nights hereafter) in which we could make 
use of 1200 s per night, and normal observation runs (NON nights 
hereafter) in which the telescope was available during the whole 
night for our project. The observational procedure was as follows:
\begin{itemize}
\item{RON nights: on dark nights one image of 1200 s was taken, 
otherwise (moon nights) several short exposures, each of 300-400 
s, were performed and then re-centered on selected field stars, and 
averaged to give the total exposure. The position of each 
individual field star was measured using the {\it imexamine} 
task and images were combined using the {\it imcombine} task.}
\item{NON nights: under photometric conditions, $BVRI$ photometry of 
QSO0957+561 was performed. Landolt standard fields (Landolt 1992) 
were observed to provide the photometric calibration. When the
nights were not of photometric quality several exposures of 1200 s 
each were performed in every filter to obtain a final deep exposure 
by averaging them.}
\end{itemize}

The final data set is composed by 15 $B$, 14 $V$, 44 $R$, 19 $I$ brightness
measurements in the 96 season; 13 $B$, 25 $V$, 72 $R$, 18 $I$ data points
in the 97 season; and 34 $B$, 33 $V$, 104 $R$, 31 $I $ data points in the 
98 season. High quality photometry in $BVRI$ was obtained on 30
nights during 1997 and 1998. Mean results for two reference stars
(H and D, see Fig. 1) and QSO components are given in Table 1. It 
is important to mention that QSOB magnitudes were corrected from
light of the lens galaxy following the procedure explained in 
\S \ref{PSFphot}.

\section{Data reduction process} \label{red}

A remarkable characteristic of the photometric data presented here 
is their high degree of homogeneity; they were obtained using the 
same telescope and instrumentation over the entire monitoring 
campaign. Therefore, the reduction process can be the same for all 
the frames. In a first step, the data were reduced using the {\it 
ccdred} package. The overscan was subtracted from the images, which 
were then flat-fielded using very high signal-to-noise master 
flats, each of them taken from the mean of ten sky flat exposures 
made shortly before the beginning of the observations. These basic 
CCD reductions (bias, flat-field) are crucial when the noise must 
be kept as low as possible. However, to attempt the observation of 
quasar brightness fluctuations of $\sim$ 0.01 magnitudes -in order 
to detect short-timescale microlensing events- a high level 
of photometric accuracy is needed. To this end it is crucial to 
separate every source of error adopting specific solutions for 
each of them. There are two main sources of error in CCD 
photometry of QSO 0957+561 system:
\begin{enumerate}
\item{{\bf Extinction errors:}
It is known that the main part of the variability of the observed 
target magnitude is explained in terms of atmospheric extinction 
and air-mass variability. Extinction errors are complicated 
by color terms when broad multi-band photometry is dealt with.} 
\item{{\bf Aperture Photometry errors:}
Due to the special configuration of QSO 0957+561 system, there are 
some specific aperture photometry errors to take into account. As
demonstrated in Colley \& Schild (1999), these errors are driven 
by seeing variations, and can be separated in two parts as follows, 
1) Influence of the lens galaxy: Since the core of the giant 
elliptical lens galaxy of $R=18.3$ is only separated by $1 \arcsec$ 
from the B image, most of the galaxy's light lies inside the image 
B aperture, but outside the image A aperture. This effect could 
introduce errors of order 1-2\% in the final measured fluxes from 
images A and B (see Colley \& Schild 1999). 2) Overlapping of 
images. The separation between the two images is $6\farcs1$ and 
hence, when poor seeing conditions prevail, there is an important 
effect of cross-contamination of light between the two quasar 
images.} 
\end{enumerate}

As explained above, the amount of archived data is so large (more 
that 1 thousand 1kx1k CCD images) that an automated photometry code 
is necessary. For extinction errors, the best and traditional 
method to work with is to measure differential photometry with several 
field stars close to the lens components (Kjeldsen \& Frandsen 
1992). 

However, the solution for aperture photometry errors presents a 
higher level of difficulty. The only automated solution offered to 
date is explained in Colley \& Schild (1999). These authors used 
HST data (Bernstein et al. 1997) of the lens galaxy for subtraction 
and reference stars to estimate the level of cross-talk between the 
images. After these corrections, they found that photometry is 
reliable to about 5.5 mmag (0.55 \%) over three consecutive nights 
of real data. In this paper an alternative solution to the problem 
is proposed: PSF fitting using DAOPHOT software. A new, completely 
automatic IRAF task, {\it pho2com}, has been developed. Using a 
sample of simulated data it is demonstrated that the proposed 
scheme can reach high precision photometry; 0.5 \% for B component 
and 0.2 \% for A component. The following two sections are devoted 
to explaining each of the adopted solutions to eliminate CCD 
photometry errors.

\subsection{PSF photometry: the {\it pho2com} IRAF task} \label{PSFphot}

It is well known that PSF fitting is the most precise method to 
carry out photometry of faint and/or crowded field stars, whereas 
aperture photometry is better for brighter, isolated stars. In 
order to benefit from these facts the {\it pho2com} task, written 
in the IRAF command language, combines aperture photometry (APPHOT 
package) and PSF fitting (DAOPHOT package) as explained below. 
Before applying the {\it pho2com} task it was necessary to select 
an image as a reference image and re-center all the frames, using 
accurate centroid determination from field stars, to the reference 
one. The {\it pho2com} photometry has two main iterations:
\begin{itemize}
\item{Iteration 1: accurate sky background determination.

A precise determination of the sky background is extremely 
important for accurate photometry. There are mainly two different 
ways to find the sky background: global-sky or local-sky 
determination. Whereas in the local-sky method the sky value is 
calculated from pixels around objects, in the global-sky 
determination the sky is described by a simple, slowly changing 
function of the position in the field, e.g., a plane. This last 
method is the most precise, but uncrowded fields are necessary in 
order to prevent sky level changes from field stars. This is the 
case of TwQSO field where most pixels see a background sky value 
unperturbed by stars, so  the global-sky option was used for sky 
determination. The main steps of current iteration are:
\begin{enumerate}
\item{Reference stars and QSO components were removed from the 
frame using PSF fittings ({\it allstar} DAOPHOT task). This was 
done, as explained above, to prevent perturbations from these 
objects in the sky determination.}
\item{The sky level was determined by means of a smooth surface 
fitting ({\it imsurf} task) to the frame. The resulting image of 
iteration 1 is a sky-subtracted frame.}
\end{enumerate}
}

\item{Iteration 2: object photometry.

As commented above, the {\it pho2com} tasks uses aperture 
photometry for reference stars and PSF fitting for TwQSO 
components. Following Stetson (1987), the PSF is defined 
from a small sample of isolated stars (G, H, E, D stars in our 
case). The PSF fit has two components: an analytic and an 
empirical one. For the 2D analytic function the user can select 
between an elliptical Gaussian, an elliptical moffat function, an 
elliptical Lorentzian and a Penny function consisting of an 
elliptical Gaussian core and Lorentzian wings. These functions
were applied to each frame, selecting the one
which yields the smaller scatter in the fit. For the PSF empirical 
component a linear variation with position in the image proved to 
give the best results. The main steps in iteration 2 are:
\begin{enumerate}
\item{Applying aperture photometry with a variable aperture of 
radius=2xFWHM (the FWHM was measured from reference stars) the 
reference stars fluxes were extracted. It is important to remember 
that the frames resulting from iteration 1 are sky-subtracted, 
and therefore the sky background value was forced to zero in the 
aperture intensity extractions.}  
\item{PSF fit photometry, with a variable aperture of radius=FWHM, 
was applied to all the objects.}
\item{Aperture corrections were computed from the previous data 
to compare the QSO component fluxes with reference stars 
(aperture correction will transform data with radius=FWHM to 
radius=2xFWHM) and standard stars (aperture correction will 
transform data with radius=FWHM to 
photometric standard star radius, normally 4xFWHM).}  
\end{enumerate}
}
\end{itemize}

A sample of simulated astronomical data was chosen in order to 
test the performance of the {\it pho2com} task. Simulations were 
made with the {\it artdata} package. Each simulated frame  
included the lens galaxy, the A and B quasar components and the 
D and H reference stars (see Table 1 for photometric data). The 
lens galaxy was created with a de Vacouleurs (elliptical) light 
distribution, $I(r)=exp\{-7.67[(r/R_e)^{(1/4)}-1]\}$ with 
$R_e=4\farcs5$, taking into account published HST data (Bernstein 
et al. 1997) and ground-based photometry (Schild \& Weekes 1984, 
Bernstein et al. 1993). The accurate position of each object was 
also defined using HST astrometry. Finally 200 simulated images 
were created with the {\it mkobjects} task. The only free 
parameter (see Table 2) was the atmospheric seeing, which was 
simulated with values between  $0\farcs9$-$2\farcs7$ (see Figs. 
\ref{simuA} and \ref{simuB}). Effects of pixellation and noise 
were included (for more details see {\it mknoise} task). Noise 
effects were considered by adding a Gaussian 
and Poisson noise to the images, which have a constant background 
(for each filter a mean sky value is deduced from real data). 
This kind of ideal photometry is not, of course, a full noise 
description. In any case the main error sources (lens galaxy 
light contamination and cross-talk between components) were 
included in the simulated images so the final estimated 
errors should be considered first order ones, where high order 
corrections (faint neighboring stars or galaxies, basic CCD 
reductions.., see Gilliland et al. 1991) are neglected. Aperture 
(with a fixed radius of $3\arcsec$) and {\it pho2com} photometry 
was applied to the simulated images. Differential light curves 
are plotted in Figs. \ref{simuA} and \ref{simuB}. Correlations 
with seeing variations are clear. Although the seeing profile is 
the same for the two reference stars and the QSO 0957+561 components, 
fixed aperture photometry has final mean errors of $\approx 
1.5\%$, $\approx 2.2\%$ for A and B components respectively. 

PSF fitting photometry improves aperture photometry magnitudes, 
but subtraction of the lens galaxy is still not perfect and some 
of its light is present in the final B component magnitude; 
therefore the final QSOB magnitudes are over-estimated. To 
correct B magnitudes from underlying galaxy light linear 
relations between seeing and magnitude errors were calculated by 
means of simulated data. Figures \ref{seeerrA} and 
\ref{seeerrB} are plots of magnitude errors for A and B components 
versus seeing for $BVRI$ filters. After correcting data for these 
errors, final errors of $\approx 0.2\%$, $\approx 0.5\%$ were 
obtained, for the A and B simulated components, respectively. Two main 
conclusions can be deduced: 1) As explained above, the B component 
presents higher errors than A, due to its proximity to the lens 
galaxy; 2) Because the lens galaxy is extremely red (Schild \& 
Weekes 1984), QSOB magnitude errors are larger in the red colors. 
Real data were also corrected for underlying galaxy light using 
the linear correlations of Figs. \ref{seeerrA} and 
\ref{seeerrB}.

\subsection{Differential photometry} \label{diffpho}

The basic technique of differential photometry is very simple, 
and consists in determining the difference, in terms of 
magnitude, of the A and B images to selected field stars. 

The transformation equations used to obtain the standard 
magnitudes are the following:

\begin{eqnarray}
b=&B+B_{0} + B_{1}(B-V) + B_{2}X \\
\nonumber
v=&V+V_{0} + V_{1}(B-V) + V_{2}X \\
\nonumber
r=&R+R_{0} + R_{1}(V-R) + R_{2}X \\
\nonumber
i=&I+I_{0} + I_{1}(R-I) + I_{2}X,
\label{transform}
\end{eqnarray} 

where $BVRI$ are the standard magnitudes; $bvri$ are the instrumental
magnitudes (i.e. $r=-2.5log[F_{r}]$, where $F_{r}$ is the object 
flux through a predefined aperture); $X$ is the airmass; and
$(B_{0},V_{0},R_{0},I_{0})$, $(B_{1},V_{1},R_{1},I_{1})$, 
$(B_{2},V_{2},R_{2},I_{2})$ are the zero-point constants, the 
color term coefficients and the extinction coefficients, 
respectively, determined from observations of 
standard stars. For a given object, the main source of magnitude 
variability can be explained in terms of atmospheric extinction 
and air mass variability. The usual way to remove this error is 
to use a comparison star observed at the same time under the same 
conditions (this is one of the main advantages of CCD 
observations). Under this assumption, the differential magnitude, 
for instance $R$, is then found as 

\begin{equation} 
r_{o}-r_{s}=(R_{o}-R_{s})+R_{1}[(V-R)_{o}-(V-R)_{s}],
\label{colR}
\end{equation}

where subindices $o$,$s$ represent the object and comparison star 
respectively. The term $R_{1}[(V-R)_{o}-(V-R)_{s}]$ is very 
important and is null only if the color term of the system is 
equal to zero, $R_{1}=0$, or the target object and the companion 
star have similar colors, $(V-R)_{o}=(V-R)_{s}$. In $BVRI$ 
photometry, color terms are not zero and, to decrease errors it 
is necessary to have similar spectra for the object and the 
comparison star. In this case it is possible to approximate
$R_{o}=R_{s}+ (r_{o}-r_{s})$.

Figure \ref{fieldR} shows the field of QSO 0957+561 in the R band 
obtained as a combination of all the individual images taken 
during the three seasons. The total equivalent exposure time is 
51.46 hours and the limiting R magnitude 25. The set of potential 
comparison stars, F, G, H, E, D, were examined differentially in 
sets of 4 versus one star. This allowed us to establish the 
stability of each comparison star. After careful analysis, only 
two stars -D and H- were selected as reference stars for 
differential photometry. Photometric errors were calculated using 
the statistical error analysis developed by Howell {\it et al.} 
(1988), which uses the rms of the differential photometry of 
comparison stars (H-D in our case) to deduce the photometric errors of 
QSO components A and B. In initial rms calculations the derived 
values are higher than expected so Eq. \ref{colR} was considered  
which, for selected reference stars, can be written as
\begin{equation} 
r_{H}-r_{D}=(R_{H}-R_{D})+R_{1} \; colVR_{H-D},
\label{colRHD}
\end{equation}
where $colVR_{H-D}=(V-R)_{H}-(V-R)_{D}$ which, taking into account 
the data in Table \ref{datacolHD}, is equal to 0.08. The color 
terms $B_{1},V_{1},R_{1},I_{1}$ are not normally expected to 
change during the course of a night, as they are due to the 
mismatch between the instrumental bandpasses and the standard 
Johnson $BVRI$ bandpasses. However, instrumental bandpasses are 
derived as the convolution of the mirror reflectivities, the 
filter transmissions, and the chip response, so significant 
changes are indeed expected in the course of a season. Under 
this assumption, Eq. \ref{colRHD} can be formulated as
\begin{equation}
r_{H}-r_{D}=(R_{H}-R_{D})+f_{R}(JD)\;colVR_{H-D},
\label{colRHDjd}
\end{equation}
where $f_{R}(JD)=R_{1}$ is a smooth function of Julian Day which 
fits the possible time changes of the color term $R_{1}$. This 
equation is demonstrated in Fig. \ref{colorR}, where we plotted 
the color term  $R_{1}$ derived from Landolt standard stars and 
the same term derived from Eq. \ref{colRHDjd} using observational 
data from reference stars H,D. The curve is a parabolic 
fitting to reference star data which has, due to error 
propagation, large errorbars ($\approx 0.1$).If it is assumed 
that parabolic fitting represents real data without noise
it is clear that the smooth 
variations in the differential light curves of reference stars 
H,D are mainly due to changes in color terms. To correct $R$ data 
of color term variations (the process is equivalent for the other 
filters) the following steps were taken: 1) from the differential 
light curve of reference stars  the $f_{R}(JD)$ 
function was calculated by means of a parabolic fitting; 2) for 
reference star data the term $f_{R}(JD)\;colVR_{H-D}$ was directly 
subtracted from $r_{H}-r_{D}$ observational data obtaining the 
differential magnitude values $R_{H}-R_{D}$; and 3) for QSO data 
it was necessary to assume mean constant values for  
$colVR_{A-D}=(V-R)_{A}-(V-R)_{D}=-0.09$ and 
$colVR_{B-D}=(V-R)_{B}-(V-R)_{D}=-0.16$, and the final corrected 
R magnitudes are
\begin{eqnarray}
R_{A}=R_{D}+(r_{A}-r_{D})+f_{R}(JD)\;colVR_{A-D}=15.163+(r_{A}-r_{D})-f_{R}(JD)0.09 \\
R_{B}=R_{D}+(r_{B}-r_{D})+f_{R}(JD)\;colVR_{B-D}=15.163+(r_{B}-r_{D})-f_{R}(JD)0.16.
\label{colcorAB}
\end{eqnarray}

For the current system, the red spectra of the D reference star and 
those of QSO components are similar, so the derived color term 
correction values  are rather small, $\approx 0.5\%$ for the $R$ and $I$ 
filters. On the contrary the QSO 0957+561 is bluer than the D star, and 
in this case color term errors become as high as $\approx 2\%$, 
$\approx 5\%$ for $V$,$B$ colors respectively. The final mean errors 
for reference stars and A,B component light curves are presented 
in Table \ref{curerrors}.

\section{$BVRI$ Light Curves} \label{curves}

The results of our monitoring program are shown in Figs. 
\ref{lightR}, \ref{lightB}, \ref{lightV}, and \ref{lightI} (the 
photometric data are available at URL http://www.iac.es/lent). In 
these figures we show the light-curves and error-bars for 
components A (black circles) and B (red squares) of Q0957+561 in 
$R$, $B$, $V$, and $I$ band, where the data for the B component are 
shifted by 425 days (the time delay estimate in this paper, see 
\S 5). Final magnitudes were calculated using the {\it pho2com} 
task and finally corrected for (1) the influence of the lens galaxy 
(see \S \ref{PSFphot}) and, (2) color term variations (see \S 
\ref{diffpho}). Note the similar behavior of the curves for both 
components (especially in Fig. \ref{lightR}, corresponding to the 
$R$ band). 

The robustness of the proposed photometry method can be assessed 
by comparing the magnitudes of the QSO 0957+561 A and B components 
deduced from monitoring light curves (averaged values) and Landolt 
standard star calibrations (see Table \ref{datacolHD}). The 
calculated values are presented in Table \ref{datacom}. The 
global agreement between both sets of magnitude values is clear.

The photometric data presented in Table \ref{datacolHD} also 
needs discussion. In principle the colors of QSOA and QSOB, 
averaged over the monitoring campaign, should be essentially the 
same if sight-line-dependent extinction is ignored. A slight 
reddening is present in component A, although the significance of 
this excess, $E(V-R)=0.07 \pm 0.08$, is questionable. In order to 
verify the significance of the previous result we have plotted, in 
Figure \ref{figdifcol}, the $V$$-$$R$ color difference between 
components A and B, with B shifted by 425 days so that the 
emission time is the same for both components over the monitoring 
campaign averaged every 20 days. The ``bluing" of component A is 
now clear and we may try to understand its origin:

1)A lens galaxy absorption effect would have produced a redder, 
and not a bluer, ($V$$-$$R$) color for image B.

2)The most likely explanation, proposed by Michalitsianos et al. 
(1997), is that the ray paths of lensed components intercept 
different regions of a galactic disk associated to the host galaxy 
of the source that is viewed pole on and situated in the quasar 
rest frame.

A preliminary analysis of the the data obtained in the $R$ filter 
has yielded an important result: component B is brighter than
component A. The $R$ data have been averaged every 10 and 20 days 
and then the B component light curve has been shifted by 425 days. 
The average difference between components A and B is $m_B - m_A = 
-0.06$ mag for both the 10- and 20-day average. Moreover, the 
averaged B/A magnification ratio is 1.06, varying between 1 and 
1.12, in perfect agreement with the results described in Press 
\& Rybicki (1998), indicating the prolongation of the 
long-timescale microlensing event during 1997 and 1998. At any rate, an 
exhaustive analysis of the long-timescale microlensing in the whole 
dataset is being conducted and will be presented in a future paper. 
This study will also include a comparison between the 
short-timescale microlensing during an epoch of calmness (96/97 
seasons) and the rapid microlensing at a relatively active (but 
non-violent) epoch (the 97/98 seasons). The consequences for the 
population of dark-matter objects in the lensing galaxy and quasar 
properties will be also discussed and put into perspective.

\section{Time Delay}

Today, the historical controversy regarding the value of the
time delay of Q0957+561A, B is almost solved. After twenty years 
of monitoring, recent data establish this value at around 420 days. 
There is, however, a small controversy between two values, 
$\Delta\tau_{\rm BA}$ = 417 days (Kundi\'c et al. 1997; Pelt et 
al. 1998a) and $\Delta\tau_{\rm BA}$ = 424 days (Pelt et al. 1996; 
Oscoz et al. 1997; Pijpers 1997; Goicoechea et al. 1999). The 
difference (one week) is irrelevant in the Hubble constant 
calculations, but it may be crucial in order to detect microlensing 
events.

One of the ``classical" ways of obtaining the time delay between 
components A and B of Q0957+561 is the computation of the $A$$-$$B$
cross-correlation (see Oscoz et al. 1997, and references therein). 
In the standard procedure, the maximum of the CCF 
(cross-correlation function) is identified with the 
time delay. However, the delay-peak generally has an irregular 
shape, and this fact causes a bias in the measurement of the time
delay between the two components of the system. In this way, two
different datasets could lead to two different estimates of the 
time delay that are in appreciable disagreement. The problem was 
considered by some authors in the past. Leh\'ar et al. (1992) made a 
parabolic fit around the maximum of the cross-correlation function,
whereas Haarsma et al. (1997) used a cubic polynomial fit to the
delay-peak. Leh\'ar et al. (1992) suggested that the 
delay-peak of the cross-correlation function should be closely 
traced by the central peak (around $\tau = 0$) of the
autocorrelation function. Moreover, other features of the
cross-correlation function around lags $\tau_1$, $\tau_2$,... will
be closely reproduced in the autocorrelation function around lags 
$\tau_1 - \Delta\tau_{\rm BA}$, $\tau_2 - \Delta\tau_{\rm BA}$,..., 
respectively.

In this paper we make use of the similarity between the discrete 
autocorrelation function (DAC) of the light curve of one of the 
components (B, for example) and the $A-B$ discrete 
cross-correlation function (DCC) to improve the estimation of the 
time delay. The same origin of the A and B curves guarantees the 
fulfilment of the relationship DCC($\tau$) $\simeq$ DAC($\tau - 
\Delta\tau_{\rm BA}$) in the absence of strong microlensing 
masking the QSO's intrinsic variability. However, several questions 
such as the impossibility of observing the system during certain 
months of the year and the necessary lack of suitable edges, 
are additional drawbacks. So, the comparison 
between the DAC and the DCC from real data should be done by 
previously selecting a ``clean" dataset, i.e., a homogeneous 
monitoring of both images during two active and clear (free from 
large gaps and microlensing) epochs separated by $\sim$420 days (the 
rough estimate of the time delay). Therefore, from the DAC and
DCC functions, one can define the following function for every 
fixed value $\theta$ (days):
\begin{equation}
\delta^2 (\theta) = \left( {1 \over N} \right) \sum_{i = 1}^N 
S_i \left[ {\rm DCC}(\tau_i) - {\rm DAC}(\tau_i - \theta)\right]^2 \, ,
\end{equation}
where $S_i = 1$ when both the DCC and DAC are defined at $\tau_i$ 
and $\tau_i - \theta$, respectively, and 0 otherwise. Equation (7) 
can be minimized to obtain $\theta_0 = \Delta\tau_{\rm BA}$, the 
most probable value for the time delay. This least squares 
comparison ($\delta^2$-test) of the auto and cross-correlation 
functions enables the time delay to be determined by comparing two 
discrete series, DCC and DAC, which should, in general, have the 
same shape.

\subsection{Simulated data}

Prior to computing the time delay from real data, the 
$\delta^2$-test was applied to some simulated datasets to verify 
its reliability when dealing with discrete and irregularly sampled 
datasets. Several sets of artificial photometric data with similar 
magnitudes, error-bars and time distribution to that of the observations 
collected at Teide Observatory were created. In this section we 
will use the same terminology as with real data; that is, the 
$y$-axis will be considered as magnitude, the $x$-axis as truncated 
JD, and the delay between both curves as time delay.

A program was developed to generate sets of dates, $x_i$, 
between 1800 and 2000 (JD), approximately, with a 
pseudo-random separation taken from a uniform distribution between 
zero and five days, obtained with the G05CAF NAG function. The time 
data were then alternately separated in two different subsets, 
corresponding to A and B light curves. A first value of 
the magnitude was then calculated with the equation $y_i = F(x_i)$, 
where $F(x_i)$ is a selected function of the dates $x_i$. The
probability of measuring a value $y$ for each $x_i$, due to several
``observational effects", is proportional to ${\rm 
e}^{[-(y-y_i)^2/2\sigma_i^2]}$, and hence characterized by
$\sigma_i$, or, equivalently, the variable $d = y - y_i$ is
distributed as ${\rm e}^{[-d^2/2\sigma_i^2]}$. A $\sigma_i$ taking 
pseudo-random values between 0.01 and 0.03, obtained with the 
G05CAF NAG function, was generated for each $x_i$. From here 
the quantities $d_i$, pseudo-random numbers obtained from a normal 
Gaussian distribution with zero mean and standard deviation 
$\sigma_i$, were calculated with the G05DDF NAG function, allowing 
them to adopt positive or negative values. Finally, 
the magnitude was generated from the equation $y_o = F(x_i) + d_i 
= y_i + d_i$, with an error-bar of $\sigma_i$. The A component was 
forced to be brighter by adding 0.1 to the magnitudes of the B 
component (although this situation is not realistic, it may be 
illustrative); moreover, 420 days were subtracted from the 
JD of the A dataset to simulate the existence of a time 
delay. The result was two pseudo-random sampled functions with  
pseudo-random noise, a true delay of 420 days, and the B 
component 0.1 mag fainter than component A. The first two 
selected functions were:
\begin{equation}
{\rm F1:} \;\; y = 17.17 + 0.5 \, {\rm e}^{-0.5 f} \, \sin (f), 
\,\,\;\; {\rm where} \;\; f = {\left( x - 1800 \right) \over 20}
\end{equation}
\begin{equation}
{\rm F2:} \;\; y = 17.2 + 0.1 \, \sin (f) \, \sin(4f),
\,\,\;\; {\rm where} \;\; f = {x \over 40}
\end{equation}
An additional function, consistent with the actual variability of
Q0957+561, was created. The raw observational data, with none of 
the modifications explained in this paper, from 97--98 seasons were
selected. The light curves were then fitted by the function
\begin{equation}
{\rm F3:} \;\; y = 17.07 -0.16 \, {\rm e}^f,
\,\,\;\; {\rm where} \;\; f={-(x - 15.8 - m)^2 \over 2 \, (10 +s)^2}  
\end{equation}
$m$ being the mean of the JD in the selected range 
and $s$ its standard deviation. The resulting simulated data show 
a lower variability to that obtained from F1 and F2.

To calculate the DAC and the DCC functions the procedure 
described in Edelson \& Krolik (1988, see also Oscoz et al. 1997) 
was followed. For two discrete data trains, $a_i$ and $b_j$, the 
formula corresponding to the DCC is
\begin{equation}
DCC(\tau) = {1 \over M} \, {\left( a_i - \bar a \right) \, \left( 
b_j - \bar b \right) \over {\sqrt{ \left( \sigma_a^2 - e_a^2 
\right) \, \left( \sigma_b^2 - e_b^2 \right)}}} \, ,
\end{equation}
averaging over the $M$ pairs for which $\tau - \alpha \leq
\Delta t_{ij} < \tau + \alpha$, $\alpha$ and $e_k$ being the 
bin semi-size and the measurement error associated with the data set 
$k$, respectively. The expression for the DAC can be obtained in a 
straightforward manner from Eq. (11), while the expression for 
$\delta^2$ is given by Eq. (7). Finally, to calculate the 
uncertainty in the estimation of the time delay a Monte Carlo 
algorithm with 1000 iterations was applied to the simulations (see 
Efron \& Tibshirany 1986).

The three simulated clean datasets are shown in Fig. 12.
Open circles correspond to the A component, while red filled 
squares correspond to the B component shifted by 420 days and with 
an offset in magnitude. As can be seen, the two first sets of 
simulated data (Fig. 12, a-b) could represent violent epochs in 
the source quasar, with episodes where the variability is as much as 
0.2--0.3 mag in only 20--30 days. The last set (Fig. 12, c)
represents an epoch with less variability than the observational 
one reported in Fig. 7. The $\delta^2$-test was applied to each 
clean dataset in three different cases: (i) DAC obtained from the 
A component; (ii) DAC obtained from the B component; and (iii) 
similar to (i) but this time with a large gap in the light curve 
B (32 days for F1, 30 days for F2, and 30 days for F3). The 
resulting values for the time delay and the corresponding error 
(1$\sigma$) in days, see Table 5, clearly indicate that the 
$\delta^2$-test offers good estimates in all the simulations, even 
considering the large error-bar generated for each point, the 
existence of ``periodic" trends, and the presence of some gaps in 
some light curves. From Table 5 one can see that the maximum 
difference between the real and the central value of the derived 
time delay is of seven days (for a relatively  inactive source) and 
the 1$\sigma$ intervals always include the true delay. An 
example of the performance of the $\delta^2$-test has been plotted in 
Fig. 13. The DAC (open circles) for the A component shifted by 420 
days versus the DCC (red squares) for F2 appear in the upper panel. 
There is a very good correspondence between both curves. Possible
values of the time delay ($\theta$) versus the associated values
$\delta^2 (\theta)$, normalized by its minimum value, have also been 
represented in the lower panel.

\subsection{Real data}

The success of the calculation of the time delay from simulated 
data, as shown in \S 5.1, made it reasonable to apply the 
$\delta^2$-test to real data. The observations, collected at Teide 
Observatory, covered three consecutive seasons (1996, 1997, and 
1998), with 220 different points in the $R$ band. Some points are 
affected by strong systematic effects and show a strong and 
simultaneous variation in both components. Once these points were 
discarded, their total number was reduced to 197. Taking into account 
the presence of two main gaps in the data---JD 2450242 to 2450347 and 
JD 2450637 to 2450729---roughly corresponding to the summer months, 
two different datasets (free from large gaps and edges) can be 
selected: DSI, corresponding to the 96--97 seasons, with 28 points 
for the A component and 27 points for the B component; and DSII, 
corresponding to the 97--98 seasons, with 44 points for the A 
component and 86 points for the B component. Both DSI and DSII 
have been represented in Fig. 14, where the B light curves have 
been shifted by 420 days and +0.06 mag. As can be seen, DSI 
corresponds to an epoch of significant calmness in the activity 
of the quasar which, together with the relatively small number of 
points, made it problematical for 
time-delay calculations. This fact was stated after some tests.
On the contrary, DSII (the 97 and 98 seasons) shows some level of 
activity (although not as strong as in A95/B96) and, moreover,
contains an appreciable number of points. Neither is there any 
clear evidence for any microlensing event, a fundamental requirement 
for selecting a clean dataset. So, DSII was finally used to perform 
time-delay calculations, i.e., DSII is our clean dataset. 

The DAC and DCC functions were obtained with the same procedure 
as in \S 5.1, taking into account that the better monitoring of 
the B light curve as compared to that for the A light curve 
(see Fig. 14) made it more suitable for the DAC calculations. The 
application of the $\delta^2$-test to the DAC and DCC curves appear 
in Fig. 15 (normalized as in \S 5.1), where the minimum of the
$\delta^2$-curve appears at 425 days, corresponding to the best 
delay. The uncertainty in our estimate of the time delay was
obtained by using a Monte Carlo algorithm. A random-number 
generator added a variable to each point of DSII to simulate the
effects of the observational errors (see Efron \& Tibshirany 1986), 
standard bootstrap samples being thereby obtained. The $\delta^2$-test 
was applied to the bootstrap samples to get the time delay in 
each case, repeating the process 10000 times, a number large 
enough for the results to be treated statistically. The use of the 
Monte Carlo algorithm led to a final value of $425\pm 4$ days 
($1\sigma$). The uncertainty with the $\delta^2$-test is better 
than the uncertainties obtained with the same clean dataset with 
other alternative methods, like the dispersion spectra and the 
discrete cross-correlation techniques, $426\pm 12$ days and 
$428\pm 9$ days, respectively (see Oscoz et al. 1997 and 
references therein). On the other hand, the $\delta^2$-test with 
the DAC obtained from the A light curve gives a time delay of 
$425\pm 5$ days. The resulting DCC (filled squares) and DAC (for the 
A component, open circles) curves are presented in Fig. 16, where 
a bin semi-size of $\alpha = $ 20 days was used. The DAC has been 
shifted by 417 (upper panel) and 425 (lower panel) days. The 
disagreement between both curves is evident in the former case. 
Our study indicates that the time delay between components A and B 
of Q0957+561 must be in the interval 420--430 days and is 
therefore slightly different from the ``standard" typical value of 
417 days.

\section{Conclusions} \label{conc}

CCD observations of the gravitational lens system Q0957+561A,B in 
the $BVRI$ bands are presented in this paper. The observations, 
taken with the 82 cm IAC-80 telescope, at Teide Observatory, Spain, 
were made from the beginning of 1996 February to 1998 July, as part 
of an on-going lens monitoring program. An alternative method to
obtain accurate multi-band CCD photometry of this object is 
presented. A new, completely automatic IRAF task, {\it pho2com}, 
has been developed. This code yields accurate photometry by
simultaneously fitting a stellar two-dimensional profile to each 
QSO component by means of DAOPHOT software. Using a sample of 
simulated data, it is demonstrated that the proposed method can 
achieve high precision photometry, 0.5 \% for B component and 
0.2 \% for A component. In this paper we show that it 
is also necessary to correct $BVRI$ photometry for color term 
variations during a season, and a possible procedure is presented. 
Although PSF fitting photometry improves aperture photometry 
errors, the subtraction of the lens galaxy is still not perfect 
and some of its light is present in the final B component magnitude, 
therefore the final QSOB magnitudes are over-estimated. To correct
 B magnitudes from underlying galaxy light, linear 
relations between seeing and magnitude errors are deduced by means 
of simulated data. A remarkable characteristic of the final 
presented light-curves is their high degree of homogeneity; they 
have been obtained using the same telescope and instrumentation 
during the three years of monitoring campaign. 

A calculation of the time delay between both components by using
a clean dataset has been performed. The resulting delay, obtained 
with a new test, the $\delta^2$-test, is of 425 $\pm$ 4 days, 
slightly higher than the value previously accepted (417 days), but 
concordant with the results obtained by Pelt et al. (1996); Oscoz 
et al. (1997); Pijpers (1997) and Goicoechea et al. (1999).

\acknowledgements
We are especially grateful to E. E. Falco for advising us on the
possible presence of strange points in our datasets and to F. Atrio 
for their help in understanding some hidden aspects in our
statistical treatment.
The authors would like to thank Dr. Jes\'us 
Jim\'enez and Dr. Francisco Garz\'on for making the telescope 
readily available to us. This work was supported by the P6/88 
project of the Instituto de Astrof\'\i sica de Canarias (IAC), 
Universidad de Cantabria funds, and DGESIC (Spain) grant
PB97-0220-C02.

%\end{document}

%
% Tables
%
\clearpage
\begin{deluxetable}{ccccc}
%\scriptsize
\tablecaption{Photometric results for reference stars and QSO components
obtained from standard calibration over 30 nights on 1997 and 1998.}
\tablewidth{0pt}
\tablehead{
\colhead{Object\tablenotemark{a}}&\colhead{V}&\colhead{B-V}&\colhead{V-R}&\colhead{R-I}}
\startdata
D & 15.601$\pm$0.009 & 0.76$\pm$0.02 & 0.437$\pm$0.008 & 0.371$\pm$0.007 \\
H & 15.139$\pm$0.008 & 0.92$\pm$0.01 & 0.520$\pm$0.007 & 0.451$\pm$0.006 \\
QSOA & 17.55$\pm$0.06 & 0.19$\pm$0.18 & 0.35$\pm$0.06 & 0.27$\pm$0.06 \\
QSOB & 17.46$\pm$0.06 & 0.17$\pm$0.18 & 0.28$\pm$0.06 & 0.27$\pm$0.06 \\
\enddata
\tablenotetext{a}{see Figure 1.}
\label{datacolHD}
\end{deluxetable}
\clearpage

\begin{deluxetable}{lc}
%\scriptsize
\tablecaption{{\it Mkobject} parameters used for generating the 
astronomical images.}
\tablewidth{0pt}
\tablehead{
\colhead{Parameter}&\colhead{Value}}
\startdata
Poisson background (ADU) & 400,650,1650,1500 (B,V,R,I)\\
PSF profile & moffat \\
Seeing radius/scale (pixels) & variable (see Figs. 2,3) \\
moffat parameter beta & 2.5 \\
moffat axial ratio (minor/major) & 1 \\
Gain (electrons/ADU) & 2 \\
Read-out noise (electrons) & 5.4 \\ 
\enddata
%\tablenotetext{a}{.}
\end{deluxetable}
\clearpage

\clearpage
\begin{deluxetable}{ccccc}
%\scriptsize
\tablecaption{Photometric errors.}
\tablewidth{0pt}
\tablehead{
\colhead{Error\tablenotemark{a}}& \colhead{B\tablenotemark{b}} &\colhead{V}&\colhead{R}&\colhead{I}}
\startdata
A light curve & 0.053 & 0.023 & 0.025 & 0.037 \\
B light curve & 0.049 & 0.021 & 0.023 & 0.031 \\
H-D light curve & 0.019 & 0.007 & 0.007 & 0.008 \\
\enddata
\tablenotetext{a}{See section 3.2 for details.}
\tablenotetext{b}{Values are expressed in magnitudes.}
\label{curerrors}
\end{deluxetable}

\clearpage
\begin{deluxetable}{ccccc}
%\scriptsize
\tablecaption{Photometric magnitudes of Q0957+561A, B derived from Landolt standard 
star calibrations and from monitoring light curves as averaged values.}
\tablewidth{0pt}
\tablehead{
\colhead{QSO}& \colhead{B\tablenotemark{a}} &\colhead{V}&\colhead{R}&\colhead{I}}
\startdata
A & 17.78 (17.74) & 17.53 (17.55) & 17.17 (17.19) & 16.91 (16.92) \\
B & 17.72 (17.64) & 17.44 (17.46) & 17.14 (17.17) & 16.89 (16.90) \\
\enddata
\tablenotetext{a}{Quoted values correspond to Landolt calibrated magnitudes.}
\label{datacom}
\end{deluxetable}

\clearpage
\begin{deluxetable}{ccc}
%\scriptsize
\tablecaption{Results of the application of a Monte Carlo algorithm 
with the $\delta^2$-test to the six simulated clean datasets.}
\tablewidth{0pt}
\tablehead{
\colhead{Function}& \colhead{Time Delay} &\colhead{Comments}}
\startdata
 & 422$\pm 2$& DAC with A data\\
F1 & 422$\pm 2$& DAC with B data\\
 & 420$\pm 3$& Gap in B\\
\hline
 & 420$\pm 1$& DAC with A data\\
F2 & 419$\pm 1$& DAC with B data\\
 & 420$\pm 1$& Gap in B\\
\hline
 & 420$\pm$ 6& DAC with A data\\
F3& 420$\pm$ 7& DAC with B data\\
 & 416$\pm$ 7& Gap in B\\
\enddata
\label{datasim}
\end{deluxetable}

%
% Figures
%

\clearpage

%
% Figure captions
%
\begin{figure}
\caption{QSO 0957+561 R field obtained as a combination of all 
the individual images for the three seasons. The total equivalent 
exposure time is 51.46 hours and the calibrated limiting R 
magnitude 25. The label shows the QSO components (A,B) and five 
field stars (F,G,H,D,E).}
\label{fieldR}
\end{figure}

\begin{figure}
\caption{R light curves for the A component
obtained from a simulated 
data sample. See text for details.}
\label{simuA}
\end{figure}

\begin{figure}
\caption{R light curves for the B component
obtained from a simulated 
data sample. See text for details.}
\label{simuB}
\end{figure}

\begin{figure}
\caption{Magnitude errors
versus seeing for the A component,  
obtained from a simulated 
data sample applying PSF
fitting photometry 
by means of pho2com IRAF task.}
\label{seeerrA}
\end{figure}

\begin{figure}
\caption{Magnitude errors
versus seeing for the B component,  
obtained from a simulated 
data sample applying  PSF fitting 
photometry by means of
pho2com IRAF task. B component 
presents higher errors than 
A, due to its proximity to the
lens galaxy.}
\label{seeerrB}
\end{figure}

\begin{figure}
\caption{R color term  
from Landolt standard stars (square points); and from 
Eq. 4 using observational data of reference 
stars H,D (circle points). The continuous line is a parabolic fitting 
to reference star data. From this figure it is possible to conclude that 
the smooth variations in the differential 
light curve of reference stars H,D are mainly  
due to changes in R color term.}
\label{colorR}
\end{figure}

\begin{figure}
\caption{Light curves for A and B 
images of QSO 0957+561 in R band 
obtained between February 1996 and July 1998. 
One-sigma error bars are indicated. See 
text for observation and reduction details.}
\label{lightR}
\end{figure}

\begin{figure}
\caption{Same as Fig. 7 for B filter.}
\label{lightB}
\end{figure}

\begin{figure}
\caption{Same as Fig. 7 for V filter.}
\label{lightV}
\end{figure}

\begin{figure}
\caption{Same as Fig. 7 for I filter.}
\label{lightI}
\end{figure}

\begin{figure}
\caption{V-R color difference between A and B (shifted by 425 days) 
component over the monitoring campaign averaged every 20 days.}
\label{figdifcol}
\end{figure}

\begin{figure}
\caption{Simulated clean datasets obtained from three different 
functions (see text). The circles correspond to the A component, 
while the red filled squares represent the B component shifted 
by 420 days and with an offset in magnitude.}
\label{}
\end{figure}

\begin{figure}
\caption{Upper panel: DAC (open circles, shifted by 420 days) 
versus DCC (red squares) for F2. Lower panel: The results of the 
$\delta^2$-test (divided by its minimum value) offer the expected 
delay, i.e., 420 days.}
\label{}
\end{figure}

\begin{figure}
\caption{Datasets for 97 (upper panel) and 98 (lower panel) 
seasons. In both cases the B component has been shifted by 420 days 
and has and offset of +0.06 mag. 
The data affected by strange CCD behavior have been removed (see 
text).}
\label{}
\end{figure}

\begin{figure}
\caption{Application of the $\delta^2$-test to the clean dataset 
DSII, normalized by its minimum value. The best time delay obtained
is 425 days.}
\label{}
\end{figure}

\begin{figure}
\caption{DAC versus DCC for the clean dataset DSII. The 
DAC has been shifted by 417 (upper panel) and 425
(lower panel) days. A time delay of 417 days is discarded.}
\label{}
\end{figure}

%\end{document}

%
% Figure draws
%

\clearpage
\begin{figure*}[h]
\epsfxsize=18 truecm
\centerline{\epsffile{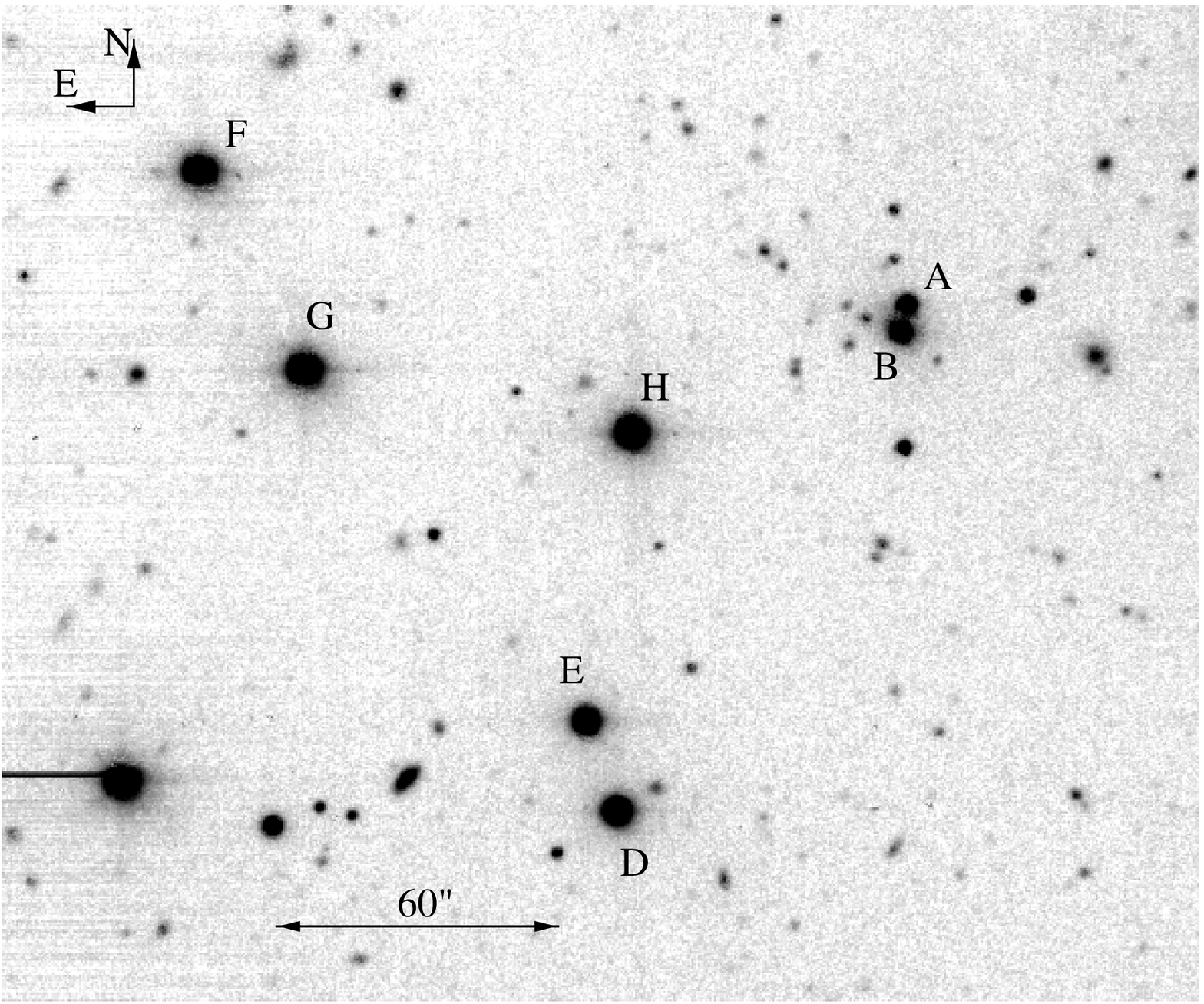}}
Fig. 1
\end{figure*}

\clearpage
.\vspace{5cm}
\begin{figure*}[h]
\epsfxsize=18 truecm
\centerline{\epsffile{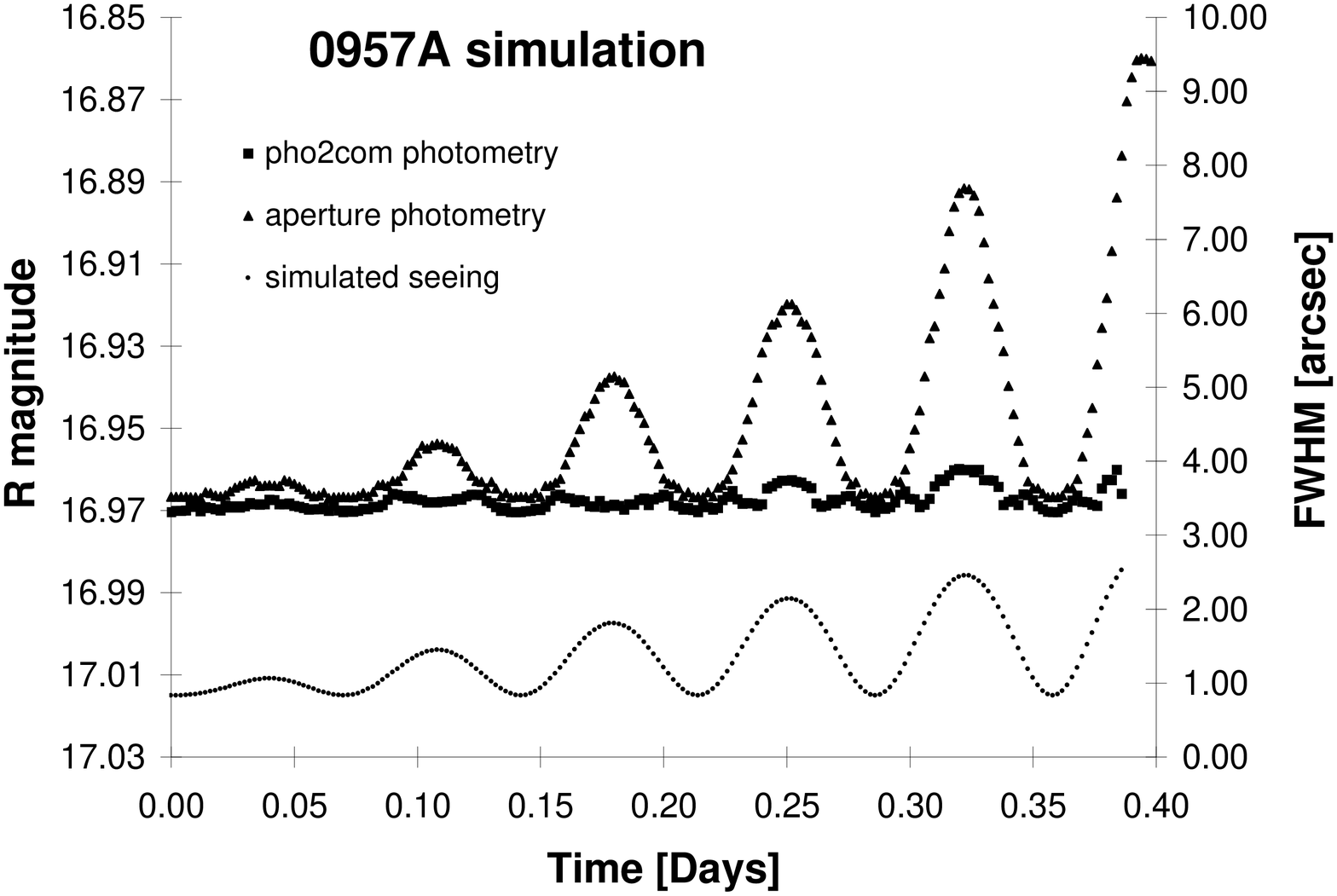}}
Fig. 2
\end{figure*}

\clearpage
.\vspace{5cm}
\begin{figure*}[h]
\epsfxsize=18 truecm
\centerline{\epsffile{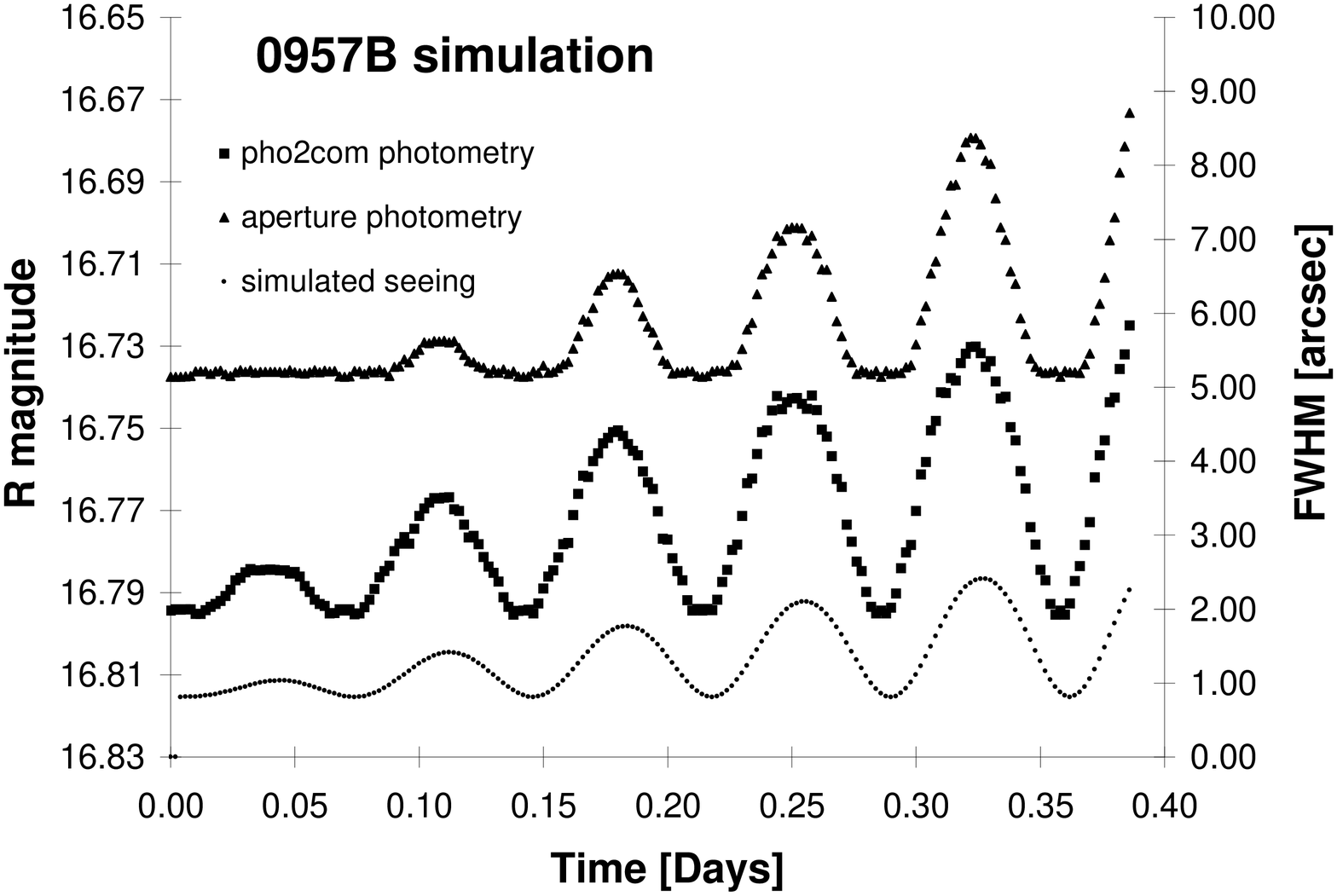}}
Fig. 3
\end{figure*}

\clearpage
.\vspace{5cm}
\begin{figure*}[h]
\epsfxsize=18 truecm
\centerline{\epsffile{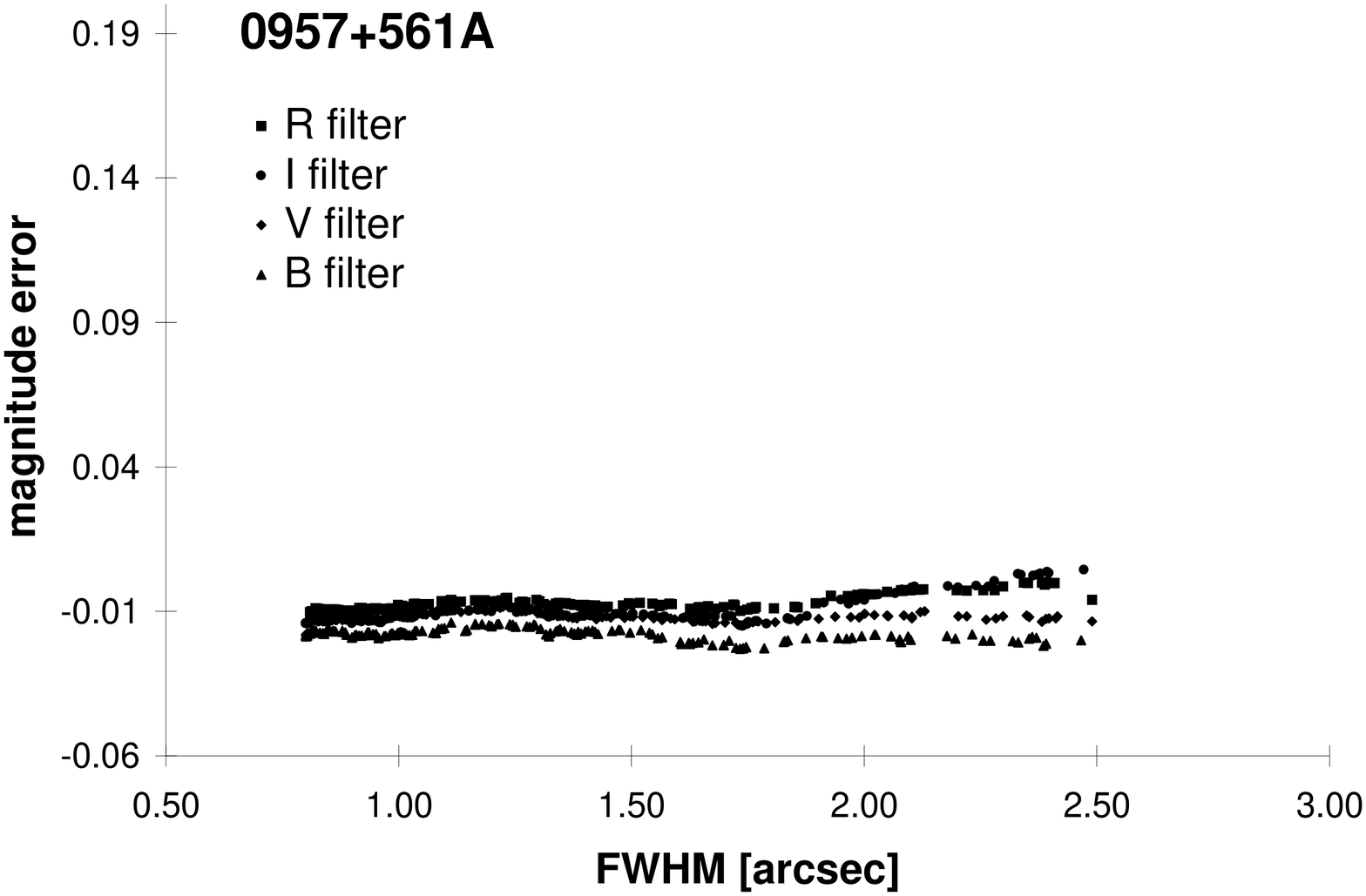}}
Fig. 4
\end{figure*}

\clearpage
.\vspace{5cm}
\begin{figure*}[h]
\epsfxsize=18 truecm
\centerline{\epsffile{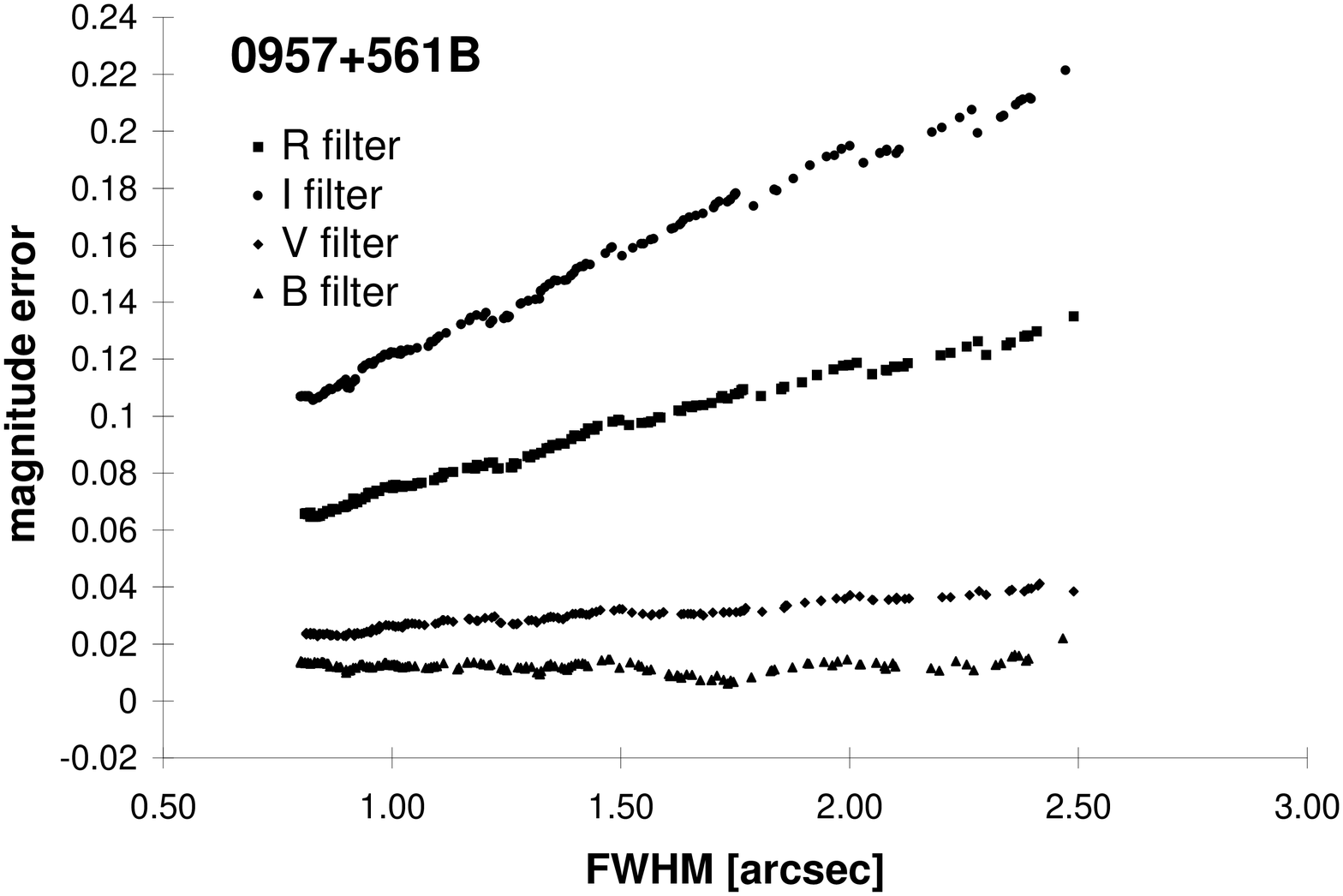}}
Fig. 5
\end{figure*}

\clearpage
.\vspace{5cm}
\begin{figure*}[h]
\epsfxsize=18 truecm
\centerline{\epsffile{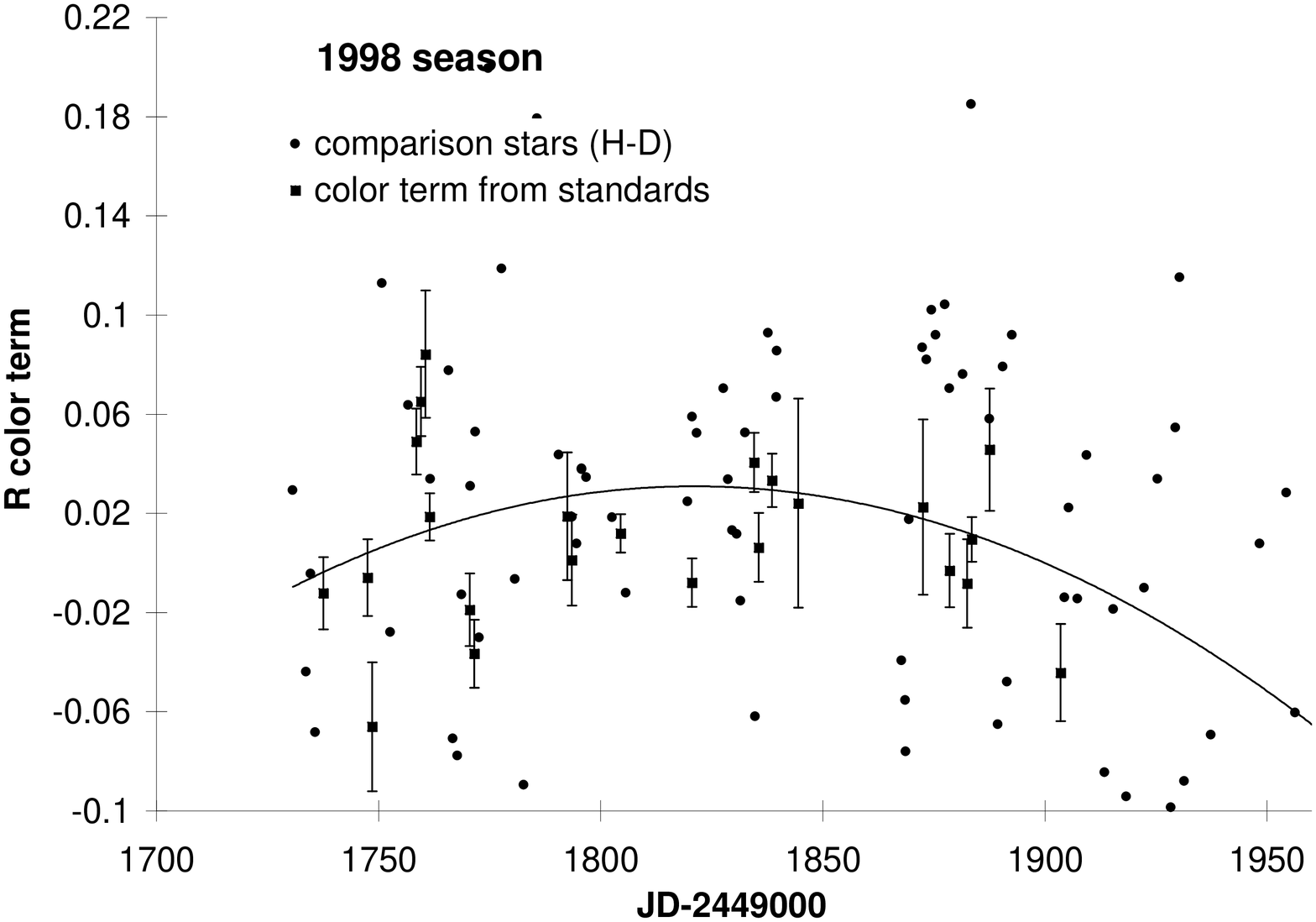}}
Fig. 6
\end{figure*}

\clearpage
.\vspace{5cm}
\begin{figure*}[h]
\epsfxsize=18 truecm
\centerline{\epsffile{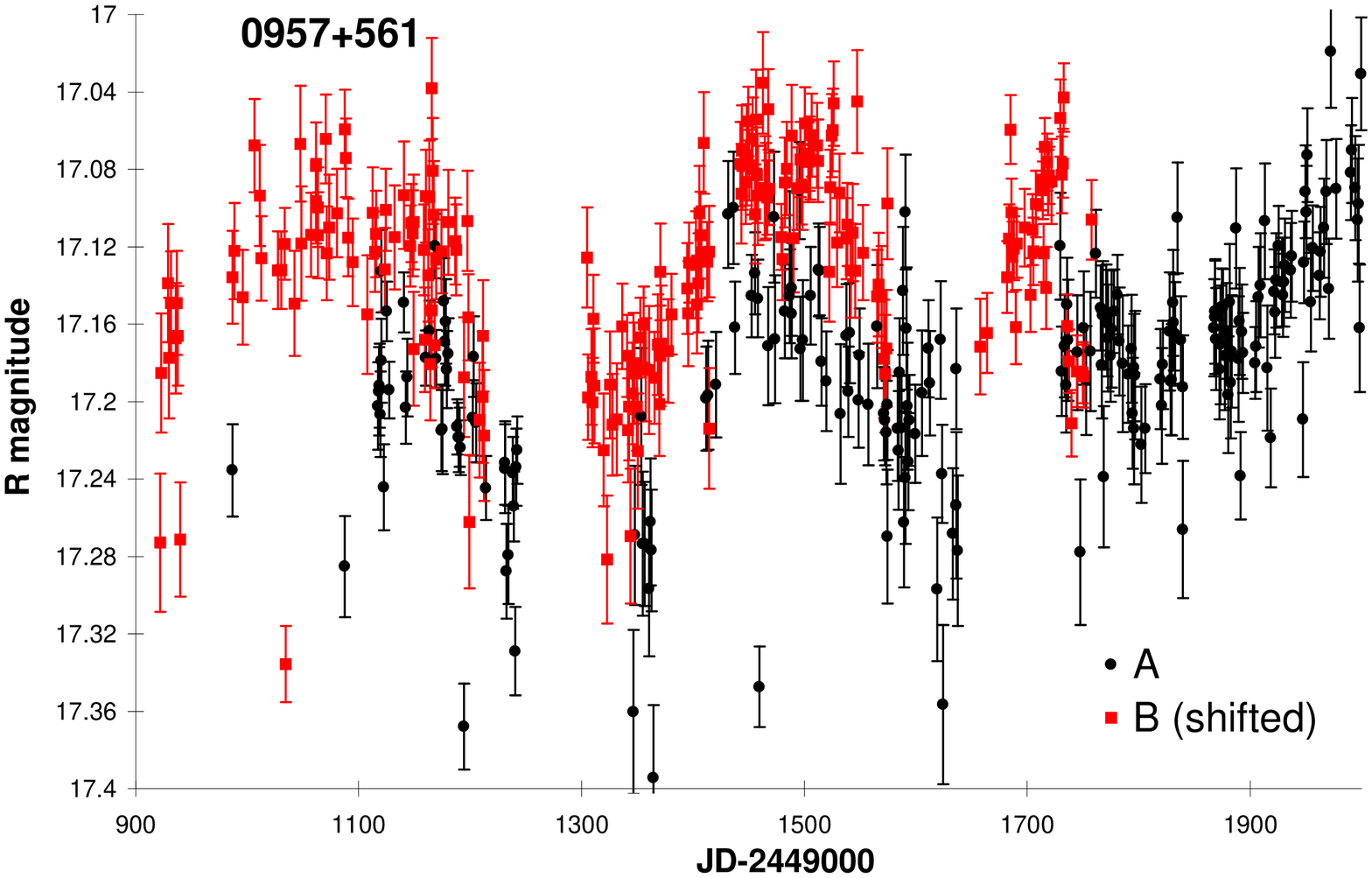}}
Fig. 7
\end{figure*}

\clearpage
.\vspace{5cm}
\begin{figure*}[h]
\epsfxsize=18 truecm
\centerline{\epsffile{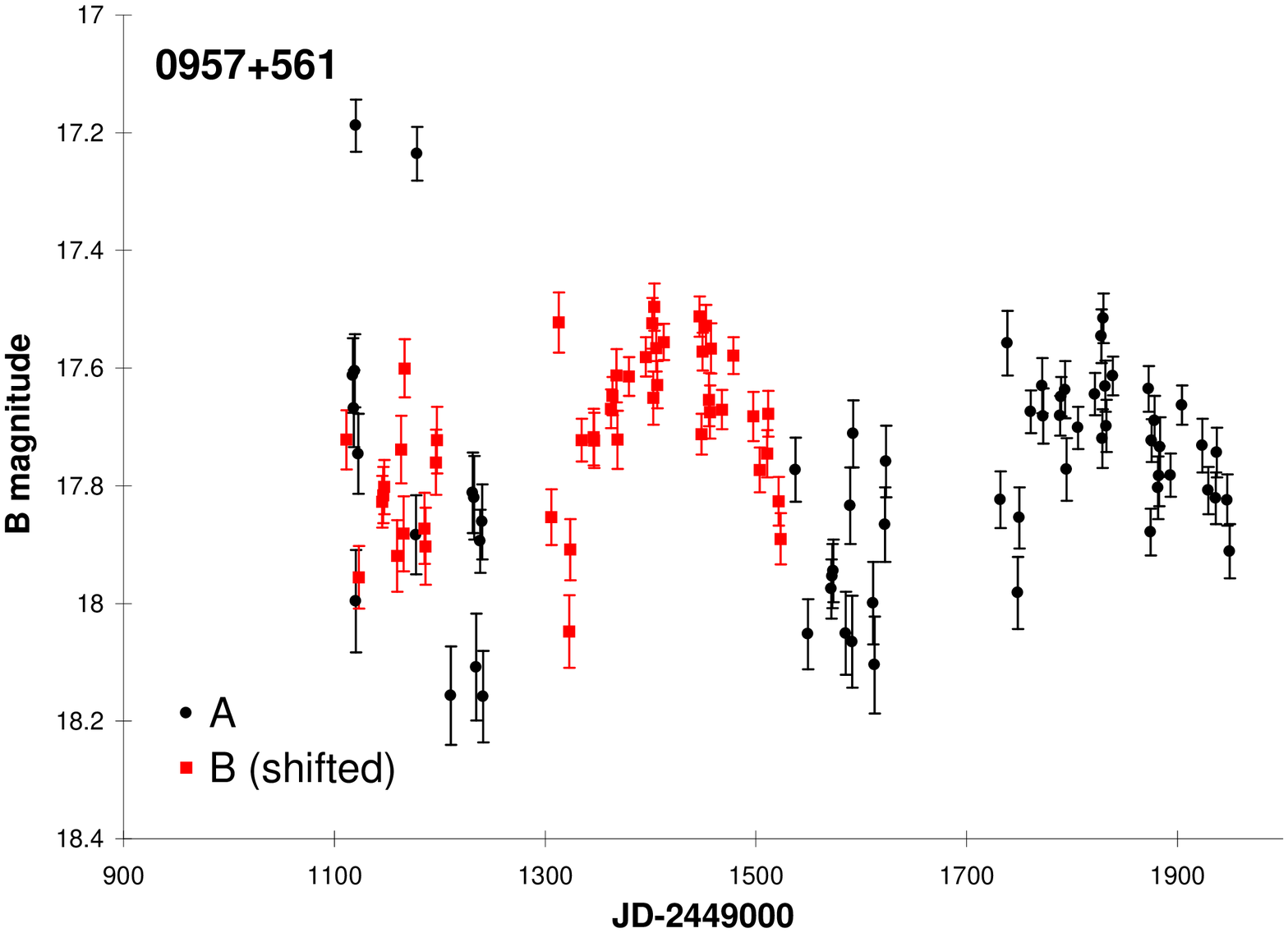}}
Fig. 8
\end{figure*}

\clearpage
.\vspace{5cm}
\begin{figure*}[h]
\epsfxsize=18 truecm
\centerline{\epsffile{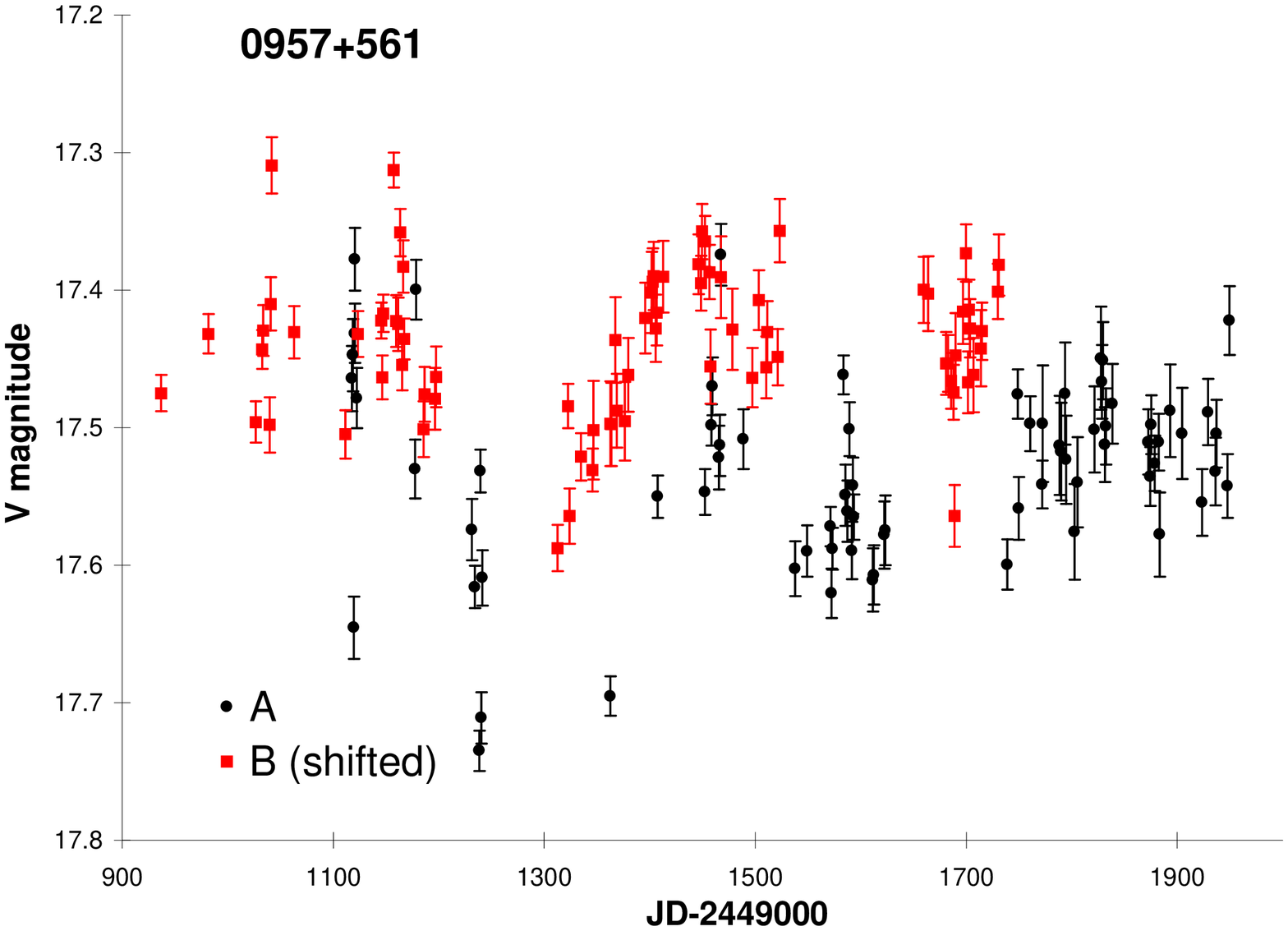}}
Fig. 9
\end{figure*}

\clearpage
.\vspace{5cm}
\begin{figure*}[h]
\epsfxsize=18 truecm
\centerline{\epsffile{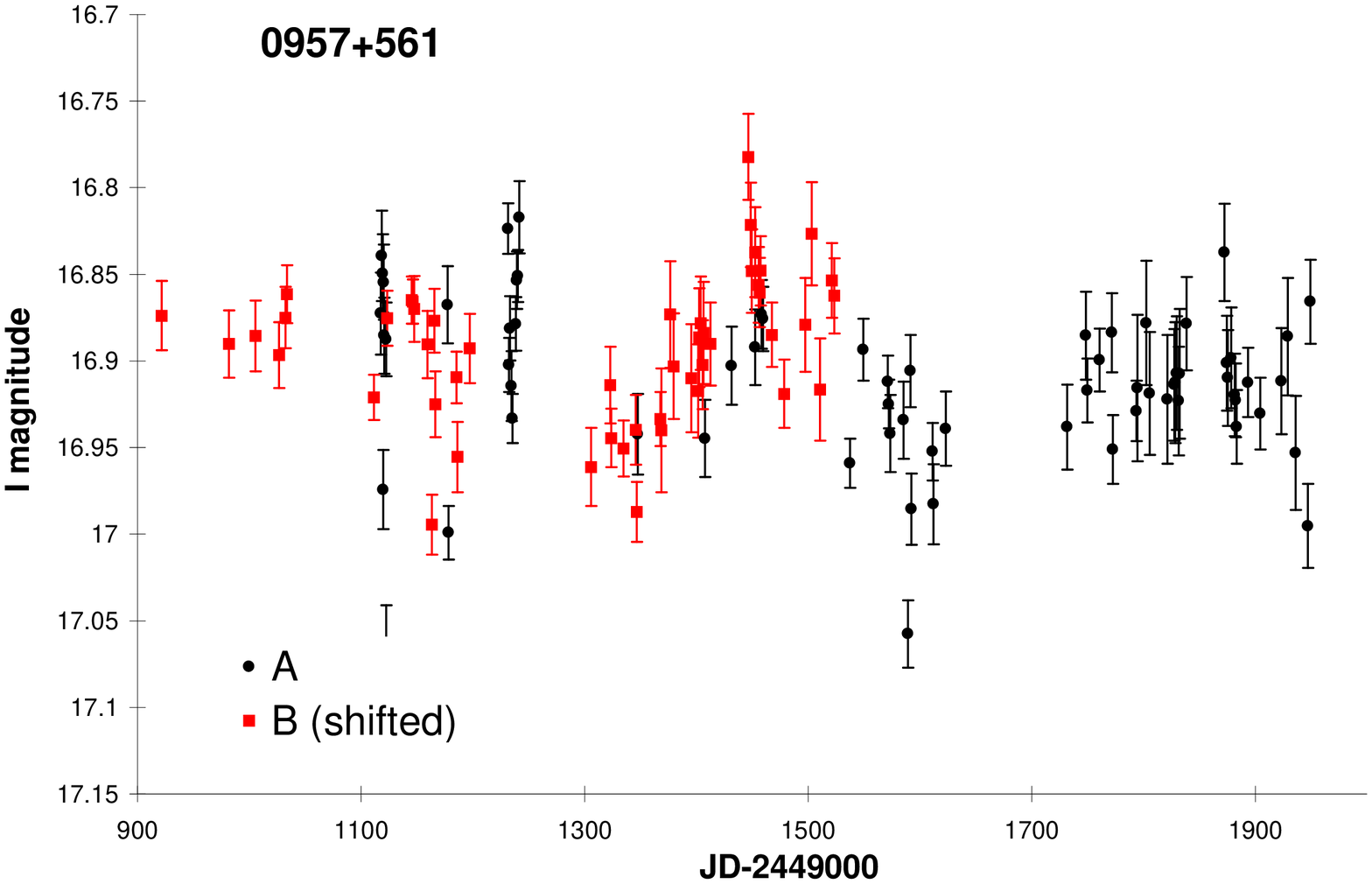}}
Fig. 10
\end{figure*}

\clearpage
\begin{figure*}[h]
\epsfxsize=18 truecm
\centerline{\epsffile{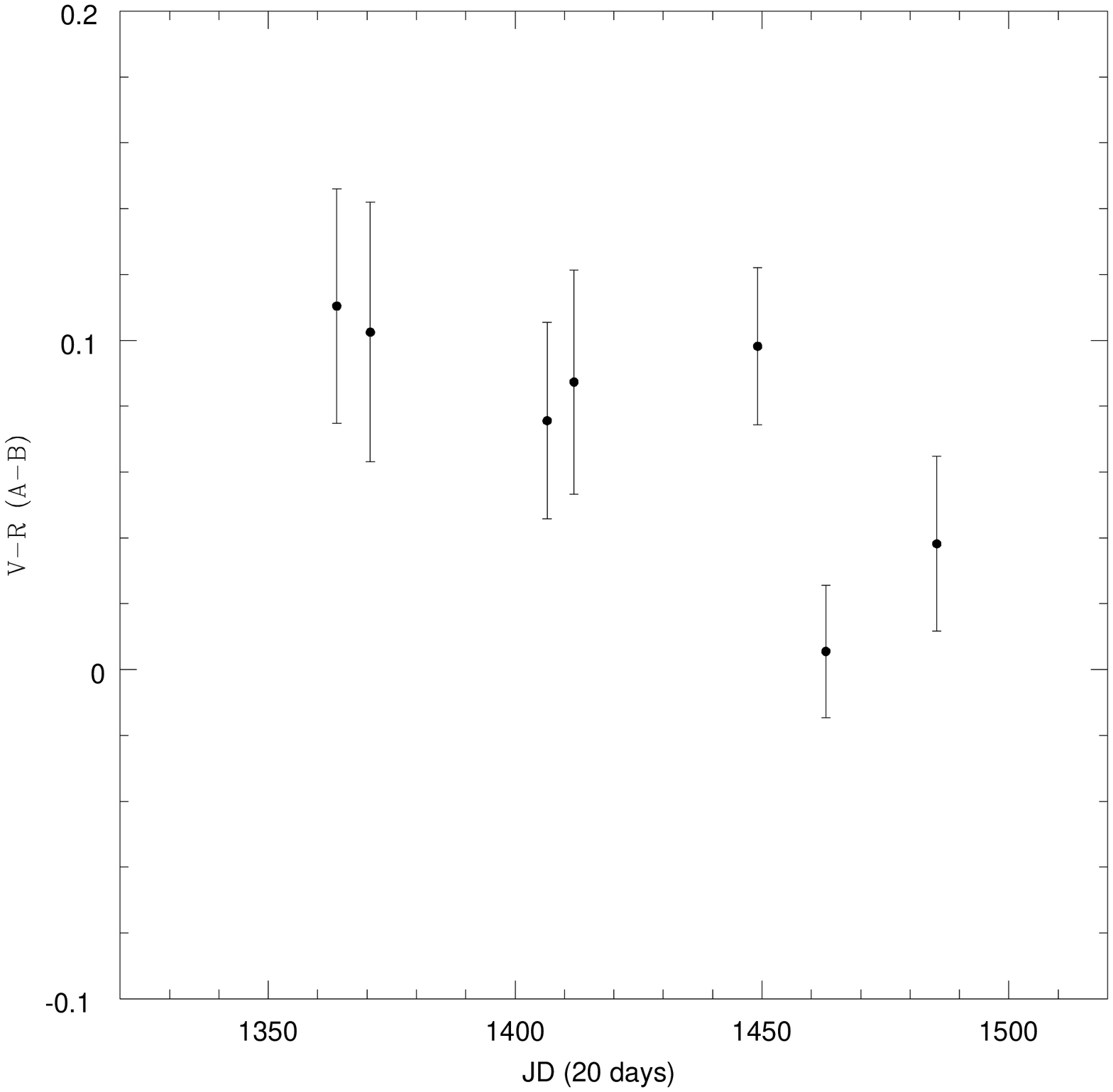}}
Fig. 11
\end{figure*}

\clearpage
\begin{figure*}[h]
\epsfxsize=18 truecm
\centerline{\epsffile{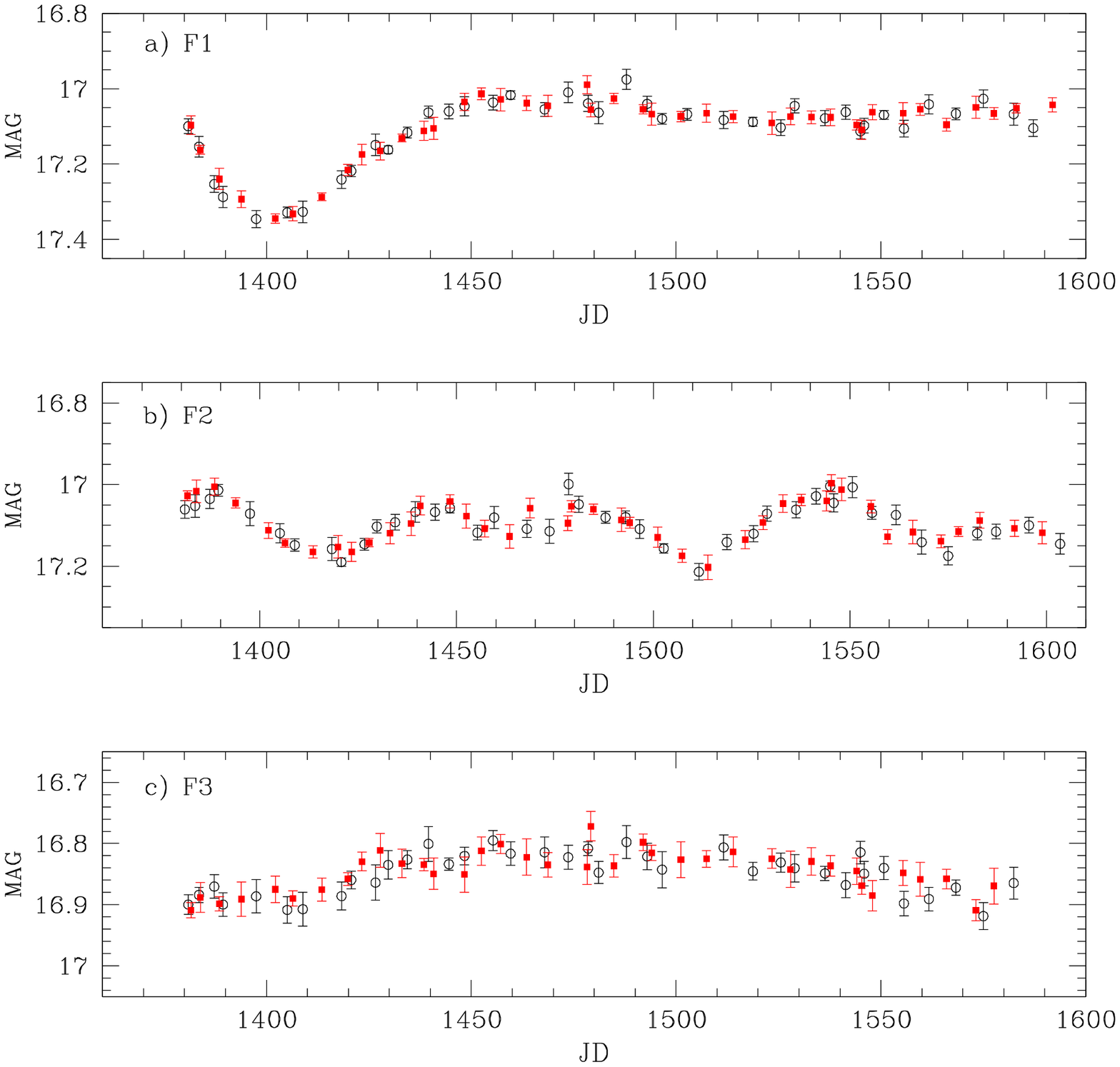}}
Fig. 12
\end{figure*}

\clearpage
\begin{figure*}[h]
\epsfxsize=18 truecm
\centerline{\epsffile{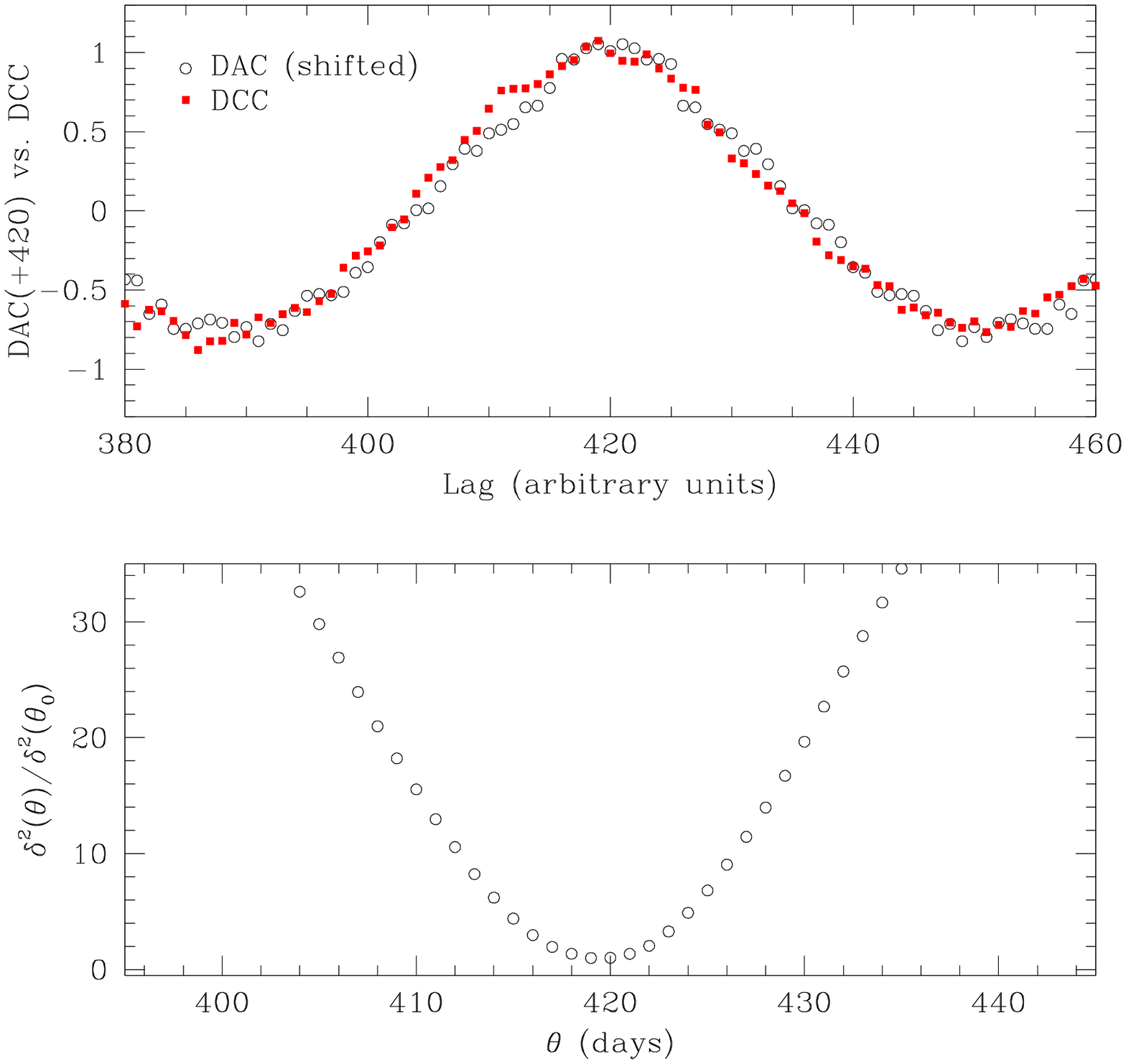}}
Fig. 13
\end{figure*}

\clearpage
\begin{figure*}[h]
\epsfxsize=18 truecm
\centerline{\epsffile{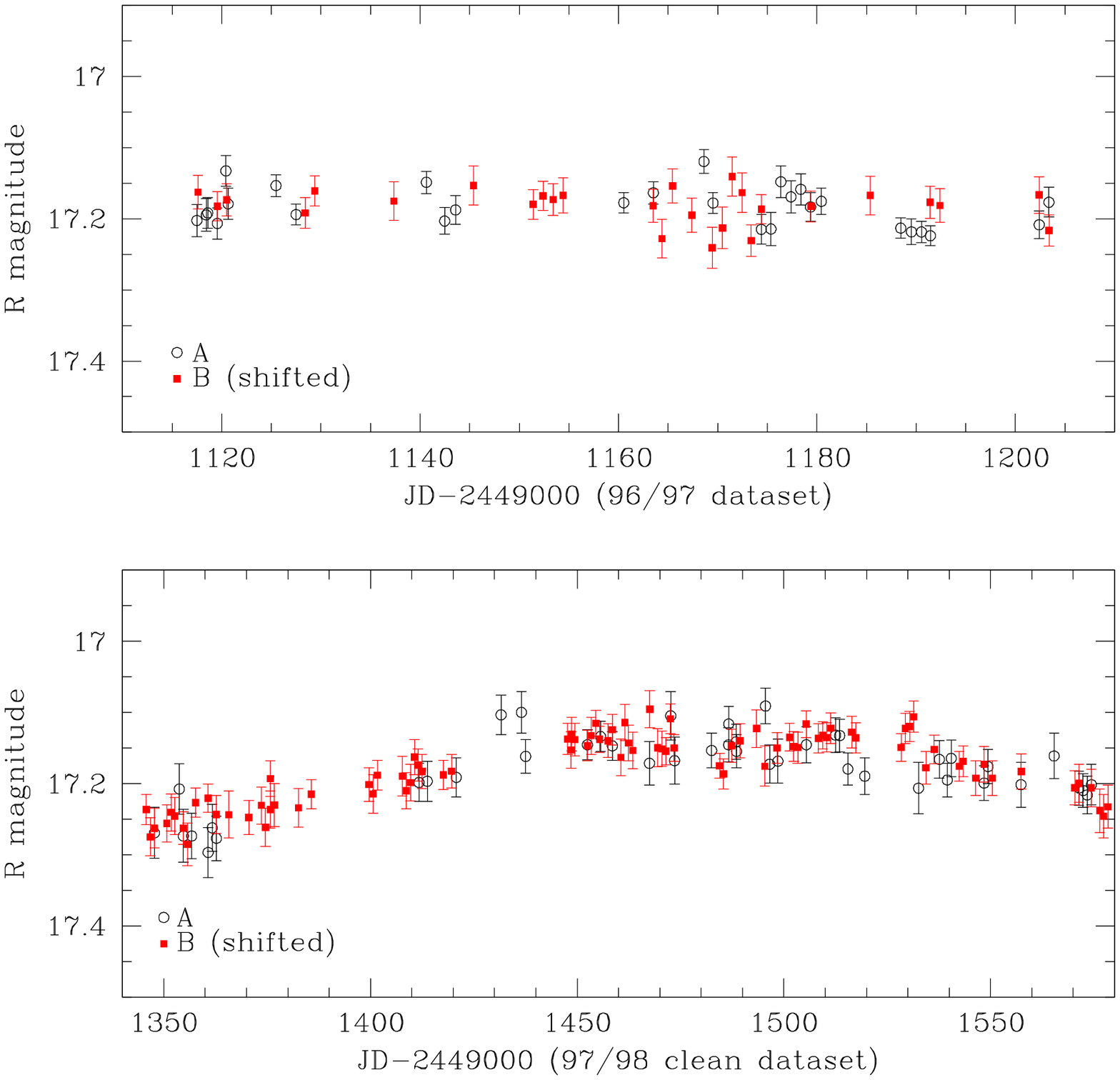}}
Fig. 14
\end{figure*}

\clearpage
\begin{figure*}[h]
\epsfxsize=18 truecm
\centerline{\epsffile{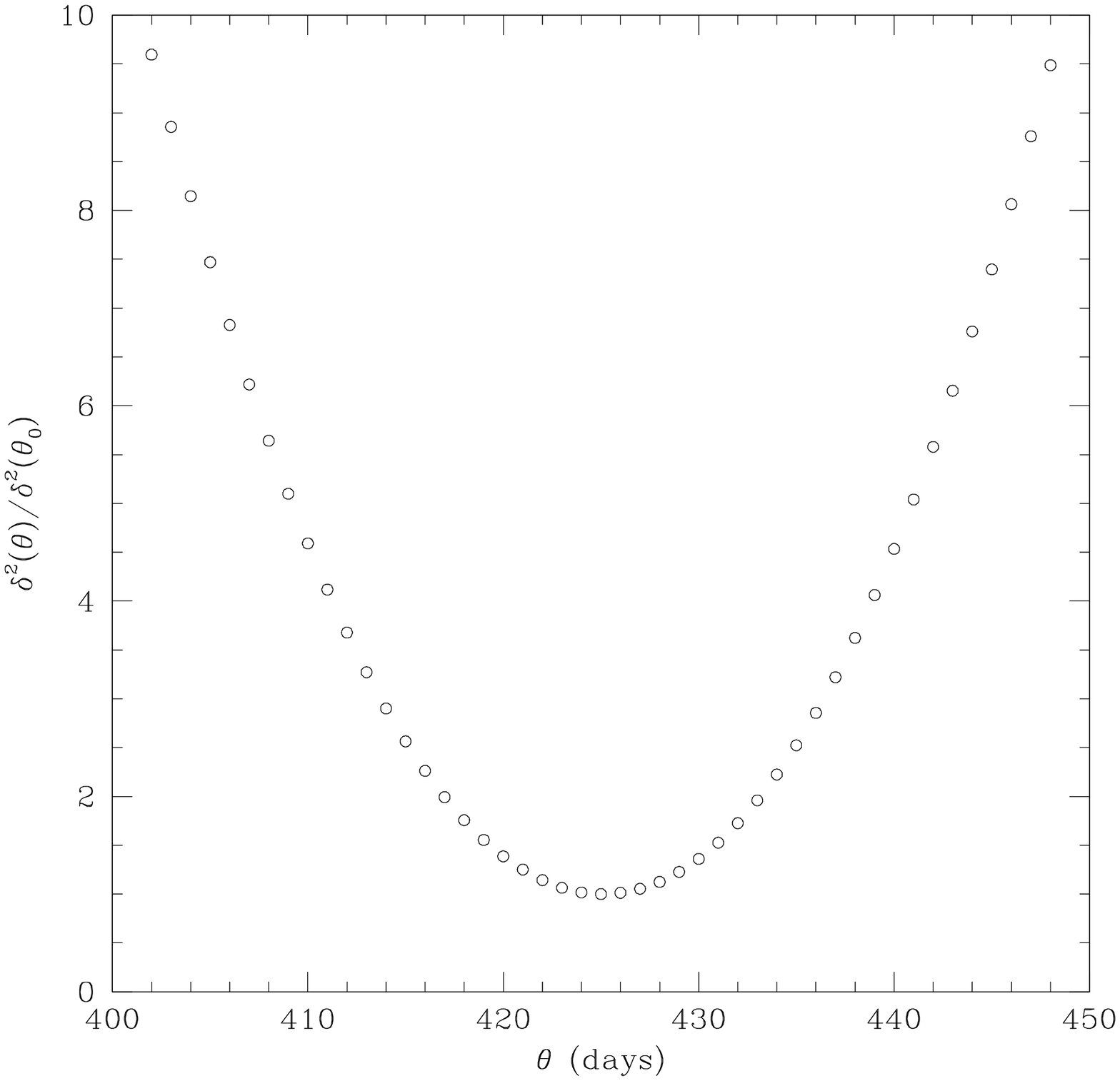}}
Fig. 15
\end{figure*}

\clearpage
\begin{figure*}[h]
\epsfxsize=18 truecm
\centerline{\epsffile{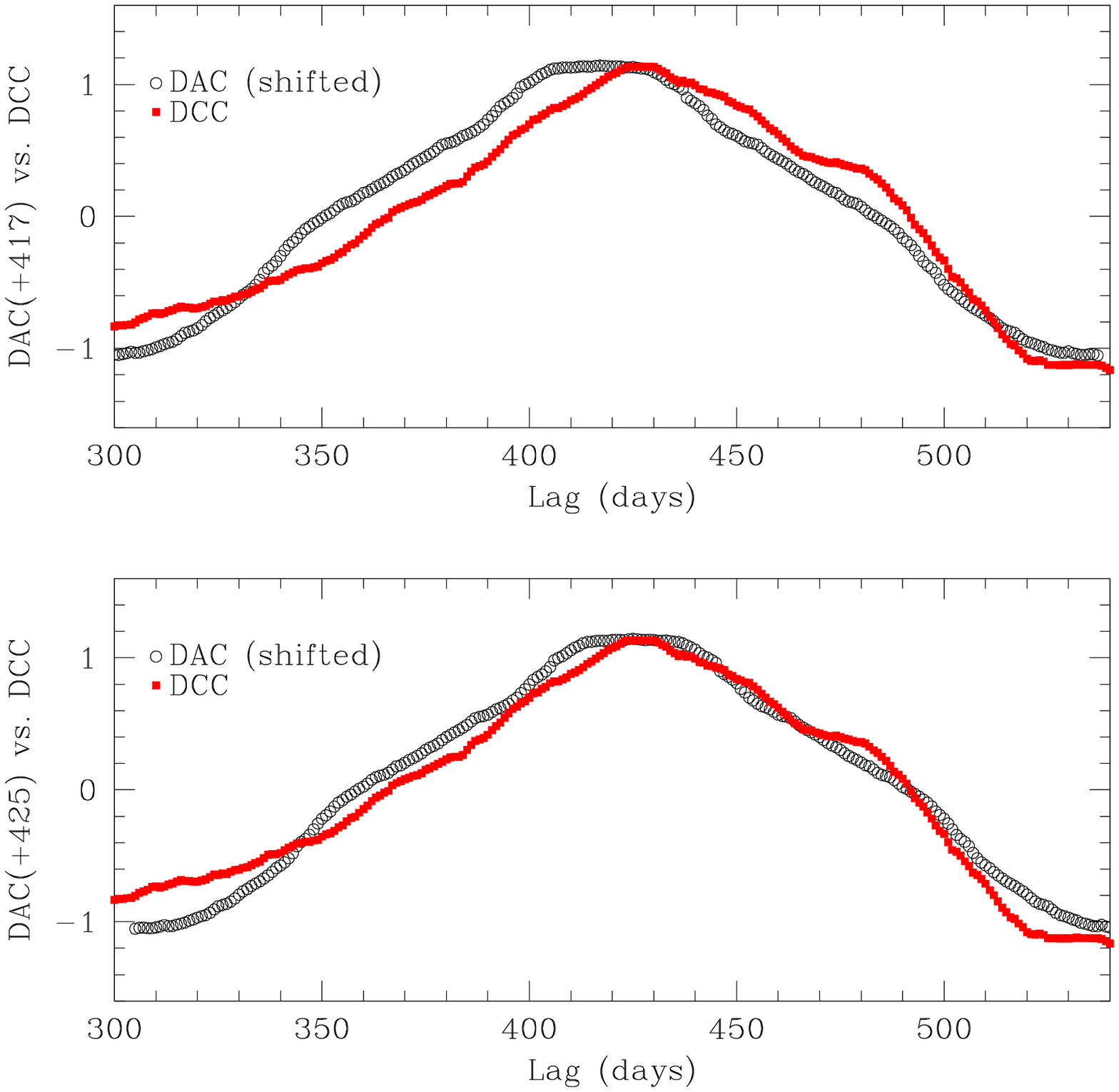}}
Fig. 16
\end{figure*}

\end{document}